\documentclass[useAMS,usenatbib]{mnras}

\usepackage{newtxtext,newtxmath}

\usepackage[T1]{fontenc}
\usepackage{ae,aecompl}

\usepackage{graphicx}	
\usepackage{amsmath}	
\usepackage{listings}

\newcommand{\teff}{$T_{\rm{eff}}$}
\newcommand{\threeD}{$\langle$3D$\rangle$}
\newcommand{\logg}{$\log{g}$}
\newcommand{\y}{$\log{\rm{(H/He)}}$}
\newcommand{\co}{CO$^5$BOLD}
\newcommand{\taur}{$\log{\tau_{\rm{R}}}$}
\newcommand{\mlt}{ML2/$\alpha$}

\newcommand{\s}{Schwarzschild}

\newcommand{\rb}{$r_{\rm{B}}$}

\title[3D DBA spectra]{3D spectroscopic analysis of helium-line white dwarfs}

\author[Cukanovaite et al.]{Elena Cukanovaite$^1$\,{\Huge \footnotemark},
Pier-Emmanuel Tremblay$^1$, Pierre Bergeron$^2$, Bernd Freytag$^3$, 
  \newauthor{Hans-G\"{u}nter Ludwig$^4$ and Matthias Steffen$^5$}
  \\
$^{1}$ Department of Physics, University of Warwick, Coventry CV4 7AL, UK \\ 
$^{2}$ D\'epartement de Physique, Universit\'e de Montr\'eal, C.P. 6128, Succ. Centre-Ville, Montr\'eal, QC H3C 3J7, Canada \\
$^{3}$ Department of Physics and Astronomy, Uppsala University, Box 516, 751 20 Uppsala, Sweden \\
$^{4}$ Zentrum f\"ur Astronomie der Universit\"at Heidelberg, Landessternwarte, K\"onigstuhl 12, 69117 Heidelberg, Germany \\
$^{5}$ Leibniz-Institut f\"ur Astrophysik Potsdam (AIP), An der Sternwarte 16, D-14482 Potsdam, Germany \\
}

\date{Accepted XXX. Received YYY; in original form ZZZ}

\pubyear{2020}

\begin{document}
\label{firstpage}
\pagerange{\pageref{firstpage}--\pageref{lastpage}}
\maketitle

\begin{abstract}
In this paper, we present corrections to the spectroscopic parameters of DB and DBA white dwarfs with $-10.0 \leq$~\y~$\leq-2.0$, $7.5 \leq$~\logg~$\leq 9.0$ and $12\,000$~K~$\lesssim$~\teff~$ \lesssim 34\,000$ K, based on 282 3D atmospheric models calculated with the \co~radiation-hydrodynamics code. These corrections arise due to a better physical treatment of convective energy transport in 3D models when compared to the previously available 1D model atmospheres. By applying the corrections to an existing SDSS sample of DB and DBA white dwarfs, we find significant corrections both for effective temperature and surface gravity. The 3D $\log g$ corrections are most significant for $T_{\rm eff} \lesssim 18,000$\,K, reaching up to $-$0.20 dex at $\log g = 8.0$. However, in this low effective temperature range, the surface gravity determined from the spectroscopic technique, can also be significantly affected by the treatment of the neutral van der Waals line broadening of helium and by non-ideal effects due to the perturbation of helium by neutral atoms. Thus, by removing uncertainties due to 1D convection, our work showcases the need for improved description of microphysics for DB and DBA model atmospheres. 
Overall, we find that our 3D spectroscopic parameters for the SDSS sample are generally in agreement with \textit{Gaia} DR2 absolute fluxes within 1-3$\sigma$ for individual white dwarfs.
By comparing our results to DA white dwarfs, we determine that the precision and accuracy of DB/DBA atmospheric models are similar. For ease of user application of the correction functions, we provide an example \textsc{Python} code.
\end{abstract}


\footnotetext{E-mail: E.Cukanovaite@warwick.ac.uk}



\section{Introduction}

White dwarfs are the most common class of stellar remnants, with around 97\% of all stars in the Milky Way destined to become this type of compact object \citep{althaus2010}. As such, they are important for studies of stellar evolution and of remnant planetary systems \citep{veras2016,vanderburg2015,manser2019,pred2019,doyle2019,vanderbosch2019,vanderburg2020}, given that all known planet-hosting stars will end their lives as white dwarfs. These stellar remnants do not undergo any nuclear burning in their cores, resulting in a relatively simple and well-understood evolution, which makes them accurate clocks for ageing different stellar populations, such as  local stars \citep{tremblay2014,fouesneau2019,fantin2019} or the inner halo of the Milky Way \citep{kalirai2012,kilic2019}.

The intense gravitational field of a white dwarf results in gravitational settling of the heaviest elements, leading to an atmosphere made up of the lightest element present \citep{schatzman48}. White dwarf classification is based on their spectral appearance, with around 80\% of white dwarfs in magnitude-limited samples having hydrogen-dominated (DA) atmospheres \citep{kleinman13,kepler2019}. The rest are referred to as non-DA white dwarfs, most of them exhibiting helium-dominated atmospheres. White dwarfs that show only He~\textsc{i} lines in their spectra are identified as DB, and are found in the effective temperature (\teff) range of 11\,000~K~$\lesssim$~\teff~$\lesssim$~40\,000~K~\citep{bergeron_db_2011}. Around 60-70\% of these white dwarfs are contaminated with traces of hydrogen and are therefore classified as DBA \citep{bergeron_db_2011,koester2015,rolland2018,genest2019a}. The origin of hydrogen in DBA stars is not yet understood, but has been proposed to be either residual \citep{macdonald1991,rolland2020} or accreted from external sources, such as remnant planetary material around the white dwarf \citep{bergeron_db_2011,koester2015,gentilefusillo2017,cunningham2020}. A sub-class of helium-atmosphere white dwarfs are DBZ or DBAZ stars which alongside helium and/or hydrogen show metal lines. For these types of white dwarfs the association with remnant planetary material is much clearer \citep{zuckerman2007,wilson2015,vanderburg2015} and is correlated with the presence of hydrogen \citep{gentilefusillo2017}.
 
It is assumed that helium-dominated atmosphere white dwarfs have lost the majority of their outer hydrogen shell either through the born-again scenario or during the final AGB thermal-pulse \citep{iben1983,straniero2003,werner2006}. Therefore, a better understanding of the helium-dominated atmosphere white dwarfs can shed light on these processes. These stellar remnants also raise important questions about their place in the overall picture of white dwarf evolution. By studying the atmospheric parameters and numbers of helium-dominated atmosphere white dwarfs (alongside other types of white dwarfs) as a function of \teff, we can pinpoint other processes that compete against gravitational settling in terms of governing the chemical composition of compact objects as they evolve. For example, the small fraction of DB white dwarfs in the range 30\,000 K < \teff~< 45\,000 K \citep{fontaine1987,eisenstein2006} is taken as evidence for helium-dominated atmosphere white dwarfs with very thin hydrogen layers transforming into DA stars at 45\,000 K through the process of hydrogen diffusing upwards to the surface \citep{fontaine1987,rolland2020,bedard2020}. Below \teff~$=30\,000$ K, DA white dwarfs with the thinnest hydrogen layers ($M_{\rm H}/M_{\rm WD} < 10^{-14})$ turn into DB or DBA stars due to the convective dilution of the thin hydrogen layer by the more massive, underlying, convective helium layer \citep{fontaine1987,macdonald1991,genest2019b}. Additionally, a large fraction of DA white dwarfs are predicted to transform into DB, DBA or DC stars due to convective mixing, which occurs below $\approx 18\,000$ K \citep{blouin2019,cunningham2020}. In this runaway process, the hydrogen convection zone reaches the underlying helium layer where it gets mixed into the more massive helium convection zone, turning the star into a helium-atmosphere white dwarf. Thus, it is important to determine accurate DB and DBA atmospheric parameters to understand these processes. 

For DB and DBA white dwarfs, there is a small systematic difference between the parameters derived using spectroscopic and photometric techniques \citep{tremblay2019,genest2019a,genest2019b}. Before the advent of \textit{Gaia} DR2 \citep{gaia2018}, the spectroscopic technique was assumed to be more precise, due to the uncertainties associated with white dwarf parallaxes. By employing the much more accurate and precise parallaxes from \textit{Gaia}, the photometric technique now rivals the precision of the spectroscopic technique. The surface gravities, \logg, (and therefore masses) of DB and DBA white dwarfs derived from photometry show a more uniform distribution as a function of \teff, compared to the \logg~distribution of the spectroscopic technique, which suggests that spectroscopic results may be subject to additional uncertainties from the underlying convection model or input microphyics \citep{tremblay2019}. Historically, the spectroscopic $\log g$ distribution of cool DB white dwarfs (\teff~$\lesssim 16\,000$ K) has been plagued by the so-called high-\logg~problem, where the spectroscopically determined values are much larger than predicted by evolutionary models and photometric colours  \citep{beauchamp1996,bergeron_db_2011,koester2015}. More recent results show that by calibrating the line broadening and eliminating very cool DB stars with weak lines and uncertain instrumental resolution, the high-\logg~problem is greatly diminished \citep{genest2019b}. The photometric technique is much less sensitive to the details of line broadening, but the absolute accuracy of the stellar parameters depends more critically on the uncertain relative flux calibration, for which DA white dwarf models are often employed \citep{narayan2019,gentile2020}.

In most studies the dominant uncertainty in the atmospheric parameters of cool DB white dwarfs (\teff~$\lesssim 16\,000$~K) is attributed to the implementation of van der Waals line broadening due to the neutral helium atom  \citep{beauchamp1996,bergeron_db_2011,koester2015}. 
The two most common implementations for this type of line broadening used in DB and DBA studies are the \cite{unsold_1955} theory, used in, for example, \citet{beauchamp1996} and \citet{bergeron_db_2011} and the modified \cite{deridder1976} treatment, used in \citet{beauchamp1996} and \citet{genest2019a,genest2019b}. \cite{beauchamp1996} showed that the modified \cite{deridder1976} version produces a much smoother distribution of spectroscopically-determined \logg~as a function of \teff. \cite{genest2019a,genest2019b} later showed that neither implementation gives a perfect agreement between the spectroscopic and the photometric techniques, or between the spectroscopic technique and the predictions of evoutionary models. However, from their samples it is clear that the modified \cite{deridder1976} treatment agrees better with \textit{Gaia} data. Either way, a more accurate implementation is needed since the \cite{deridder1976} version of the line broadening has been altered by \cite{beauchamp1996} to agree better with observations.

Additionally, there is the issue of non-ideal effects due to the neutral helium atom, which also become significant for \teff~$\lesssim 16\,000$ K. The current implementation used for white dwarf atmosphere models is the \cite{hummer1988} model, which depends on a free parameter, \rb, that determines the radius of the hydrogen or helium atom as a fraction of atomic radius according to the Bohr model. The commonly utilized value is 0.5 and it has been calibrated based on DA white dwarf spectra, specifically the line profiles of the higher hydrogen Balmer lines \citep{bergeron1988,bergeron1991}. A discussion on the effect of \rb~on \logg~can be found in \cite{tremblay2010}. This free parameter can potentially be adjusted to obtain a smoother \logg~distribution for DB and DBA white dwarfs.

The treatment of convective energy transport in atmospheric models of DB and DBA white dwarfs is another source of uncertainty influencing the spectroscopic parameters. In 1D white dwarf atmosphere modelling the \mlt~version \citep{tassoul1990} of the Mixing Length Theory \citep[MLT;][]{bohm1958} is employed. This theory relies on a free parameter called the mixing length parameter, \mlt, which for DB/DBA white dwarfs has been determined to be 1.25 from a comparison between the atmospheric parameters (\logg~and \teff) derived from UV and optical spectra \citep{beauchamp_1999_v777_dba,bergeron_db_2011}. \cite{tremblay_2013_spectra} showed that it is precisely the shortcomings in the MLT theory that cause a similar high-\logg~problem for DA white dwarfs (where the spectroscopically-determined \logg~are larger than predicted by evolutionary models and determined by photometry). They showed that this problem can be solved with the help of 3D radiation-hydrodynamical models, which treat convection from first principles and do not depend on any free parameters. \cite{myprecious} calculated the first 3D DB atmospheric models and found that while a single value of \mlt~= 1.25 can reproduce reasonably well the temperature distribution and UV fluxes of DB white dwarfs, no single \mlt~value can mimic the 3D spectra below \teff~$\approx$ 18\,000\,K, resulting in strong 3D $\log g$ corrections\footnote{See \cite{mysunandstars} for an alternative \mlt~calibration relevant for the size of the convection zone.}. Nevertheless, \cite{myprecious} found that 3D $\log g$ corrections do not result in obviously more accurate stellar parameters.
\cite{tremblay2019} cemented this by showing that 1D and 3D DB models provide spectroscopic parallaxes (calculated from spectroscopically-determined values of \teff~and \logg, and observed magnitude) that are in similar agreement with \textit{Gaia} parallaxes. It was postulated by \cite{myprecious} that inclusion of traces of hydrogen in their 3D models could potentially lead to a better agreement with \textit{Gaia}. However, given the known issues with the microphysics of line broadening in cool DB and DBA white dwarfs and concerns with the photometric calibration \citep{tremblay2019,jesus2018}, it is unclear if \textit{Gaia} can provide an accurate test of 3D convection.

In this paper we propose to use the new 3D DBA models of \cite{mysunandstars} alongside 3D DB models already presented in \cite{myprecious} to finalise the determination of the atmospheric parameters of DB and DBA white dwarfs with our accurate treatment of convective energy transport. We first introduce our 3D and reference 1D models in Sect.~\ref{sec:num}. The 3D spectroscopic corrections are determined in Sect.~\ref{sec:corr} and we apply them to observations in Sect.~\ref{observations}. In that section we also investigate van der Waals broadening and non-ideal effects and we conclude in Sect.~\ref{sec:conclusions}.

\section{Numerical setup}~\label{sec:num}

\subsection{3D atmospheric models}

We computed 282 3D DB and DBA models using the \co~radiation-hydrodynamics code \citep{freytag2012_cobold,freytag2013,freytag2017}. The models have already been presented in \cite{myprecious} and \cite{mysunandstars}, therefore we only briefly describe the model atmospheres and focus on the new spectral synthesis done in this work. Our 3D grid of models covers the hydrogen-to-helium abundance, \y, of $-10.0 \leq$~\y~$\leq -2.0$. Models at \y~$=-10.0$ are the same as the pure-helium models discussed in \cite{myprecious} and this hydrogen abundance is used for pure-helium atmosphere models since all known DB white dwarfs have upper limits on hydrogen larger than this value. Including even less hydrogen in the calculations makes no meaningful difference to the predictions. The grid also spans $7.5$~dex~$\leq$~\logg~$\leq 9.0$ dex in steps of 0.5 dex, and $12\,000$~K~$\lesssim$~\teff~$\lesssim 34\,000$ K in steps of around 2\,000 K. We show the exact values of the atmospheric parameters in Fig.~\ref{fig:models}. Additional data on the models can be found in Appendix~1 of \cite{mysunandstars}.

\begin{figure*}
	\includegraphics[width=1.5\columnwidth]{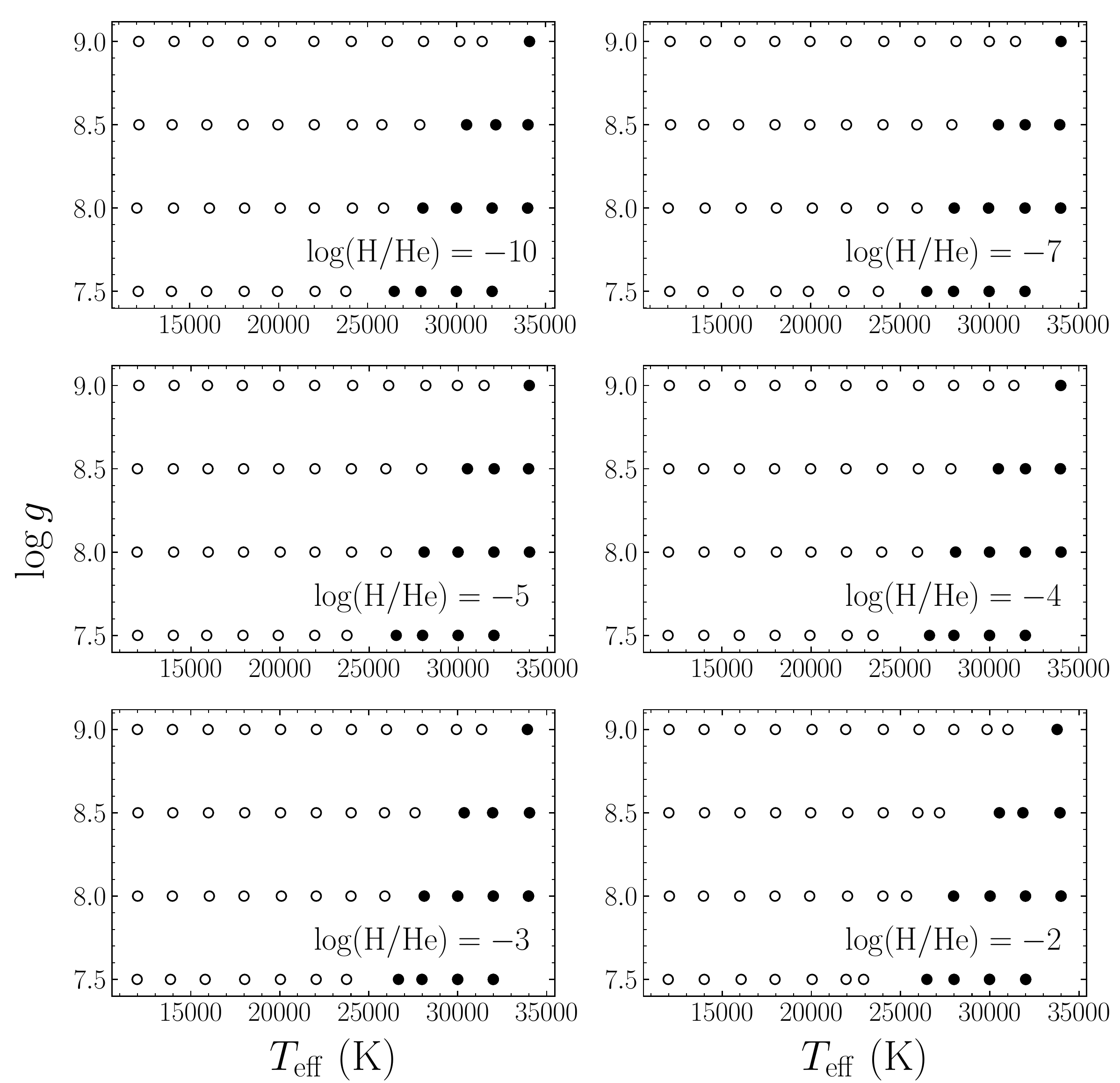}
    \caption{The atmospheric parameters of 3D DB and DBA models averaged over time and over contours of constant \taur. 3D simulations with open and closed bottom boundaries are indicated as open and filled circles, respectively. The hydrogen abundance of the models is indicated on each individual panel. } 
    \label{fig:models}
\end{figure*}

The input parameters of \co~include boundary conditions, \logg, an equation of state (EOS) and an opacity table. The \teff~is calculated only after the model has finished its run. The EOS and opacity table define the hydrogen abundance of DBA simulations. The opacity tables are binned and more details on the binning procedure can be found in \cite{nordlund_1982_opac_binning}, \cite{ludwig_1994_op_binning}, \cite{vogler_2004_op_binning} and \cite{myprecious}.
We use 10 bins with boundaries at \taur~$=$~[99.0, 0.25, 0.0, $-$0.25, $-$0.5, $-$1.0, $-$1.5, $-$2.0, $-$3.0, $-$4.0, $-$5.0] based on reference 1D models of \cite{bergeron_db_2011}. As discussed in \cite{myprecious}, due to interpolation issues we do not include the \taur~$=$~[$-$5.0,$-$99.0] bin.

Each model is run using the box-in-a-star setup of \co~\citep{freytag2012_cobold}, where a portion of an atmosphere is modelled in a Cartesian box made up of $150 \times 150 \times 150$ grid points. The side boundaries are periodic and horizontally we make sure that at least $4 \times 4$ convective granules are included. The grid spacing in the vertical ($z$) direction is non-equidistant. The top boundary of the simulation box is open both to material and radiative outflows. For all models this boundary is higher than \taur~$=-5.0$, such that the top of the photosphere is contained within the model.

The bottom boundary for all models is around \taur~$=3.0$, but in some cases the models had to be extended deeper to include the negative convective flux region found below the \s~boundary of the convection zone \citep{mysunandstars}. In those cases, the vertical extent of the box was also increased, resulting in some of the models being increased up to 250 grid points vertically. We use two types of bottom boundary prescriptions depending on the size of the convection zone \citep{freytag2017}. At the highest \teff~of our 3D grid, the convection zones become small enough to be fully vertically simulated. In those cases, we use the closed bottom boundary, which means that the bottom of the simulation is closed to material inflows, but is open to radiative flux flows. Additionally, the vertical velocity at this boundary is enforced to be zero, which we make sure is physical for any given closed bottom model. The effective temperature for these models is set by changing the value of the inflowing radiative flux at the bottom boundary. For the majority of our 3D models, however, we use the open bottom boundary, which is open both to material and radiative flux inflows. In order for this boundary to be realistic, we have to make sure we simulate enough of the convection zone such that the inflowing material at the boundary can be described by adiabatic convection. In this case, \teff~is set by specifying the entropy of the inflowing material. There is some evidence which suggests that a layer within two pressure scale heights, $H_{\mathrm{p}}$, can be affected by the boundaries of the 3D simulation \citep{Grimm_Strele_2pressurescale}. Therefore, we ensure that both the top and the bottom of the photosphere is at least two $H_{\mathrm{p}}$ away from either top or bottom boundary of the simulation. The top and bottom of the photosphere is determined based on the minimum and maximum optical depth at which the plasma becomes optically thin for photons of frequency $\nu$, i.e. \taur($\tau_{\nu} = 1$) (see \citealt{myprecious} for more information).

In order to be able to derive meaningful information from 3D models, we ensure that each model is relaxed in the second half of the run. We determine relaxation by monitoring the \teff~of the models as a function of time, making sure the fluctuations are below 1\% level. Similarly, we track the fluctuations of total flux at all depths and convergence of the velocity field as a function of run time. We spatially average the pressure, $P$, and temperature to the power of four, $T^4$, over contours of constant optical depth. We also average them temporally using more than 12 snapshots over the last quarter of the simulation. We refer to these models as \threeD~and we discuss errors associated with such averaging in Sect.~\ref{sec:threeD}.

\subsection{1D atmospheric models}~\label{sec:oneD}

In order to calculate 3D spectroscopic corrections, we use a differential fitting approach between 3D and reference 1D synthetic spectra. The EOS and opacity tables for the 3D models were calculated using the white dwarf atmosphere code of \cite{bergeron_db_2011}, referred to as ATMO in the following. However, we use the 1D LHD code \citep{caffau_2007_lhd} for determining the 3D corrections. This is because both the 3D \co~code and 1D LHD code treat microphysics (EOS and opacity tables) in the same fashion, whereas 1D LHD and 1D ATMO treat convective energy transport in terms of the \mlt~approximation~\citep{myprecious}. Therefore, by using the 1D LHD code to calculate the 3D corrections, any corrections arising from microphysics issues (such as the opacity binning) are largely eliminated, and thus only the 3D corrections arising from the treatment of convective energy transport are recovered. 

Fig. 12 of \cite{myprecious} shows that the differences between DB ATMO and LHD models are due to the binning procedure used in opacity tables. ATMO uses 1745 individual frequencies when computing the opacities, whereas for both 3D and LHD models we only use 10 opacity bins. As the number of bins increases, \cite{myprecious} found that the agreement between DB LHD and ATMO models gets better. We performed the same test for the \y~$=-2$ grid and found similar results. Thus,  as a precaution, we utilise the 1D LHD models for calculating the 3D DBA spectroscopic corrections, given that 1D LHD models were used by \cite{myprecious} to derive the 3D DB corrections.

Our 1D LHD grid spans a parameter space slightly extended compared to that of our 3D models. It covers $-10.0 \le$~\y~$\le -2.0$, $7.0 \le$~\logg~$\le 9.5$ and $10\,000 \le$~\teff~$\le 40\,000$ K. The models are in LTE and use \mlt~= 1.25, as well as the same EOS and opacity tables as those used in 3D models, which include the physics described in \cite{bergeron_db_2011} and \cite{genest2019a,genest2019b}. 

To compute the spectra for 1D LHD and averaged 3D (see below) structures, we use the 1D ATMO code. This is because neither \co~nor LHD is capable of calculating synthetic spectra. We utilise ATMO consistently to calculate spectra for \co~and LHD using the exact same numerical setup apart from the input temperature and pressure stratification. In terms of van der Waals broadening. we use the \cite{unsold_1955} treatment, unless otherwise specified, such as in Sect.~\ref{sec:micro}. We have tested and confirmed that the particular choice of line broadening theory does not impact the final 3D corrections if the line broadening is used consistently in both 1D and 3D models.

\subsection{3D synthetic spectra}~\label{sec:threeD}

Ideally, one would calculate a synthetic spectrum from a 3D atmospheric model using a 3D spectral synthesis code such as Linfor3D \citep{ludwig08}. This way all of the information from a given 3D simulation would be used, including the horizontal fluctuations. However, this process is time-consuming and typically limited to a small portion of a spectrum, e.g. a few atomic lines. 
Instead, to calculate synthetic spectra of DB and DBA white dwarfs from 3D atmospheric models, we average the models spatially and temporally as explained in Sect.~\ref{sec:num} to calculate the so-called \threeD~structure, which we then feed into the ATMO code to calculate a \threeD~synthetic spectrum. This completely neglects any horizontal fluctuations in the 3D models. For DA white dwarfs it was shown that synthetic spectra (H$\beta$ line) derived from 3D and \threeD~structures were identical within 1\% level \citep{tremblay_2013_spectra}. However, for extremely low mass DA white dwarfs, the differences could reach a few per cent \citep{tremblay2015_elm}. 
For a discussion on why 3D and \threeD~synthetic spectra can agree in some cases and disagree in others see  \cite{tremblay_2013_spectra} and \cite{tremblay2015_elm}.

To test whether the horizontal fluctuations have any effect on the derived 3D spectroscopic corrections, \cite{myprecious} computed 1.5D spectra \citep{steffen1955}. This type of spectra is calculated by assuming that each column in the 3D simulation box is an individual plane-parallel 1D model atmosphere, and for each of these atmospheres a separate spectrum is calculated. These individual spectra are then averaged together to calculate the final 1.5D spectrum. They are also averaged over three different snapshots in time. In the 1.5D method the horizontal fluctuations are enhanced \citep{tremblay2015_elm} compared to a 3D synthetic spectrum which connects nearby grid points through inclined light rays. The 1.5D and \threeD~spectra represent two extremes of combining grid points in a 3D model, such that these two types of spectra encompass a given 3D synthetic spectrum \citep{tremblay_2013_spectra}. 

For DB white dwarfs, 1.5D spectra were found to be identical to \threeD~spectra within the observational errors and thus \threeD~spectra were used for final 3D DB spectroscopic corrections. We performed the same test for DBA models with \y~$=-2.0$ and \logg~$=8.0$ and found that the corrections derived using either type of spectra gave the same results. Therefore, this agrees with the conclusions reached for 3D DB corrections, with the difference between \threeD~and 1.5D for DBA models being even smaller, resembling the results of 3D DA models. Therefore, we use \threeD~synthetic spectra when calculating 3D DBA corrections.

\section{3D DBA corrections}\label{sec:corr}

\subsection{Fitting code}~\label{sec:fitting_code}

In order to determine the 3D corrections, we want to find a 1D LHD synthetic spectrum that best matches a given \threeD~spectrum. To do this we wrote a code that fits a \threeD~synthetic spectrum with a grid of 1D LHD synthetic spectra. We define the 3D spectroscopic corrections as 
\begin{equation}
x_{\rm{correction}} = x_{\langle \rm{3D} \rangle \rm{\ value}} - x_{\rm{1D \ LHD \ fit}},
\end{equation}
where $x$ can be \y, \logg~or \teff. The code fits the optical part of the spectrum, namely the wavelength range $3500 \leq \lambda \leq 7200$~\AA. This is the same range used by \cite{bergeron_db_2011} for fitting observations and in \cite{myprecious} for 3D DB corrections. All spectra are normalised by dividing the flux at all wavelengths by the flux value at 5500 \AA, a wavelength at which there are no helium or hydrogen lines.  

The code first fits for \teff~and \logg~assuming a value of \y. The initial value of the \y~parameter does not matter, but the code converges faster if the \y~is set equal to the 3D model abundance. Once \teff~and \logg~are found, the spectrum is then fitted for \y~at fixed values of \teff~and \logg, found in the previous step. This procedure is then repeated until convergence of 0.1\% is achieved across all three parameters. If the hydrogen lines are not visible or insignificant then we do not fit for \y. This happens mostly for \y$=-10.0$ and $-7.0$ models, as well as models with higher hydrogen abundances but large \teff~values. 

Using this fitting code, we can recover the 3D DB spectroscopic corrections of \cite{myprecious}, which relied on the \cite{bergeron_db_2011} fitting code. The corrections agree within typical observational errors, except for three models where the \teff~corrections agree within 2-3$\sigma$.

\subsection{Line cores}

\cite{tremblay_2013_3dmodels,tremblay_2013_spectra} showed that convective overshoot cools the upper layers (\taur~$<-2.0$) of 3D DA models, causing them to deviate significantly from their 1D counterparts. The cores of H$\alpha$ and H$\beta$ appear too deep in 3D when compared to observations. It was found that the discrepancy is unlikely to be a numerical, structure averaging or microphysics issues with 3D DA models. \citet{tremblay_2013_spectra} chose to remove the line cores from their fitting when determining 3D DA corrections. In contrast, \cite{myprecious} determined that removal of line cores did not impact 3D DB corrections. This is because helium lines are not formed as high-up in the atmosphere as the hydrogen lines, despite a similar strength for convective overshoot. However, hydrogen lines do appear in the spectra of DBA white dwarfs and therefore we review the properties of the line cores in this section.

\begin{figure*}
	\includegraphics[width=1.5\columnwidth]{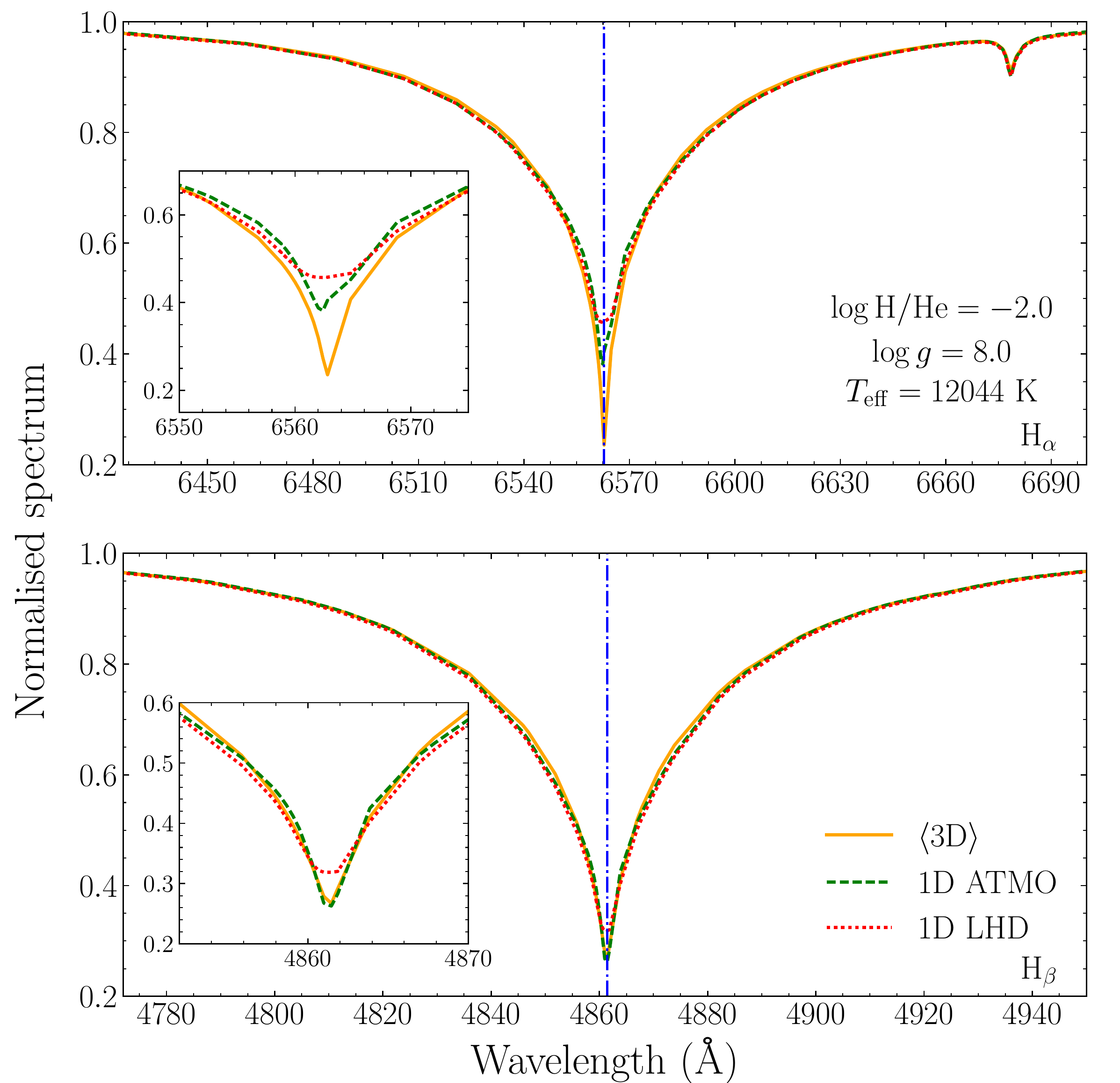}
    \caption{A comparison between the synthetic H$_{\alpha}$ and H$_{\beta}$ lines calculated from 3D, 1D ATMO and 1D LHD models for a DBA white dwarf with \y~$=-2.0$, \logg~$=8.0$ and \teff~$=12\,044$ K. The \threeD~synthetic lines are shown in solid orange, 1D ATMO in dashed green and 1D LHD in dotted red. The center of a given line is indicated by a vertical dot-dashed blue line. }
    \label{fig:line_cores_atmo_lhd}
\end{figure*}

In Fig.~\ref{fig:line_cores_atmo_lhd} we compare the line cores between 3D, 1D LHD and 1D ATMO synthetic spectra at \teff~= 12\,000\,K, $\log g = 8.0$ and \y~$=-2.0$. We chose this particular hydrogen abundance as it is the highest in our grid and thus we assume that the line core issue will be most apparent.
The 3D line core is deeper for H$\alpha$ but the difference is less pronounced for H$\beta$ when compared to DA models at the same temperature.
We also find that the 1D LHD synthetic spectrum has shallower cores than 1D ATMO. 
Therefore it appears that
the EOS and opacity tables contribute to a significant uncertainty on the prediction of the line cores, but with an effect in the opposite direction compared to 3D convective overshoot.

\begin{figure*}
	\includegraphics[width=1.5\columnwidth]{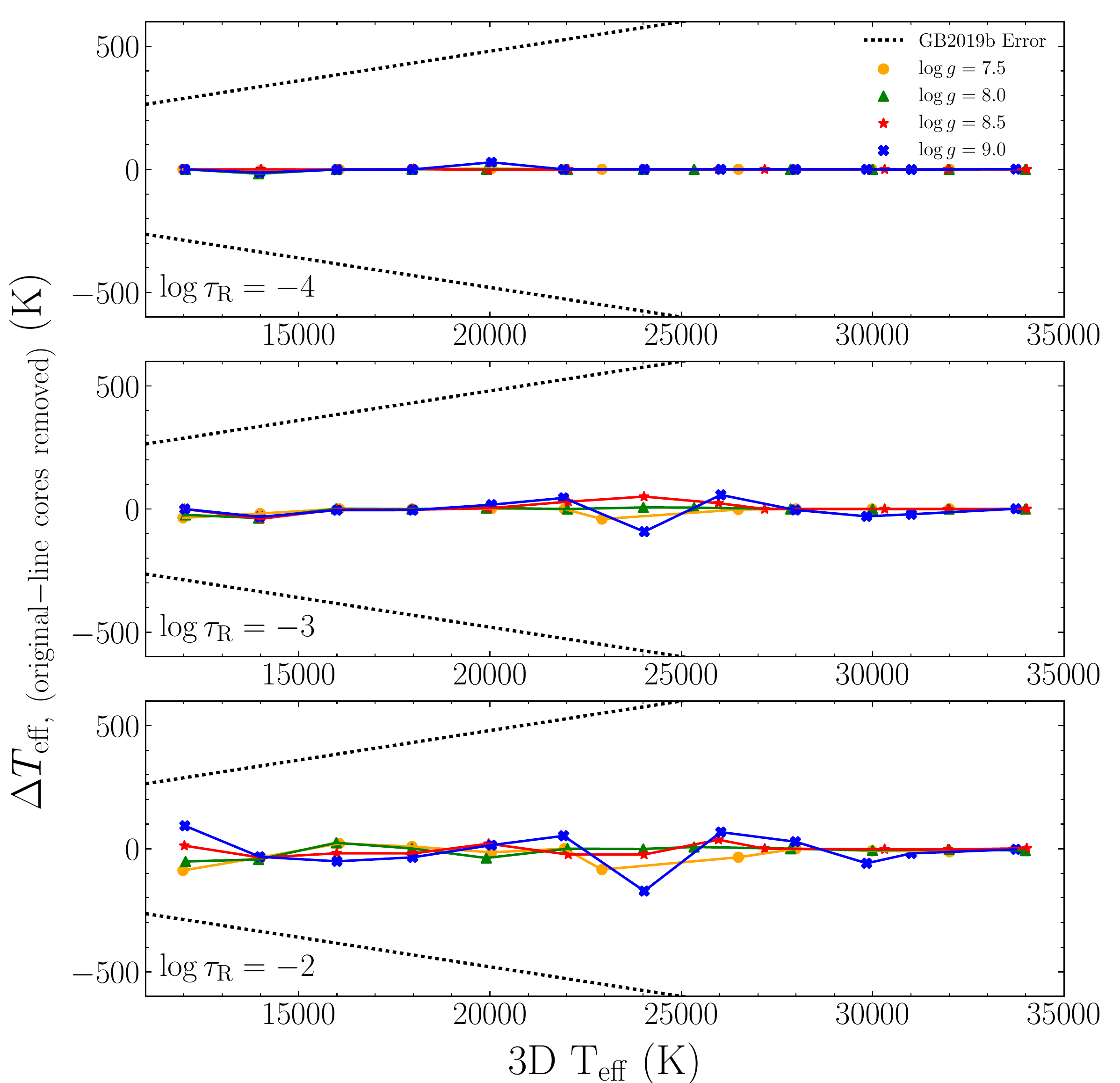}
    \caption{The difference between \teff~corrections derived from original spectra and from spectra with line cores removed for 3D DBA models with \y~$=-2.0$. The \taur~values indicated on each panel denote the atmospheric layer above which we remove any flux formed. The difference in corrections for \logg~$=7.5$, 8.0, 8.5 and 9.0 models are shown as orange circles, green triangles, red stars and blue diagonal crosses, respectively. Corrections for each \logg~value are joined for clarity. The average errors from \protect\cite{genest2019b} are shown in dotted black.} 
    \label{fig:line_cores_removal}
\end{figure*}

In Fig.~\ref{fig:line_cores_removal} we compare the 3D DBA \teff~corrections for \y~$=-2.0$ 
derived when fitting the spectrum with and without line cores. We remove line cores by removing any wavelength range corresponding to flux that was formed above a given value of \taur. As shown in Fig.~\ref{fig:line_cores_removal} the values of \taur~$=$ [$-$2.0, $-$3.0, $-$4.0] are tested. At low \teff, this will mostly remove the cores of hydrogen lines, as helium lines are formed lower in the atmosphere than \taur~$=-2.0$, but as the \teff~increases the cores of the helium lines will also be removed. We find that the removal of line cores does not affect the 3D corrections (not just \teff, but also \y~and \logg). 

\begin{figure*}
	\includegraphics[width=2\columnwidth]{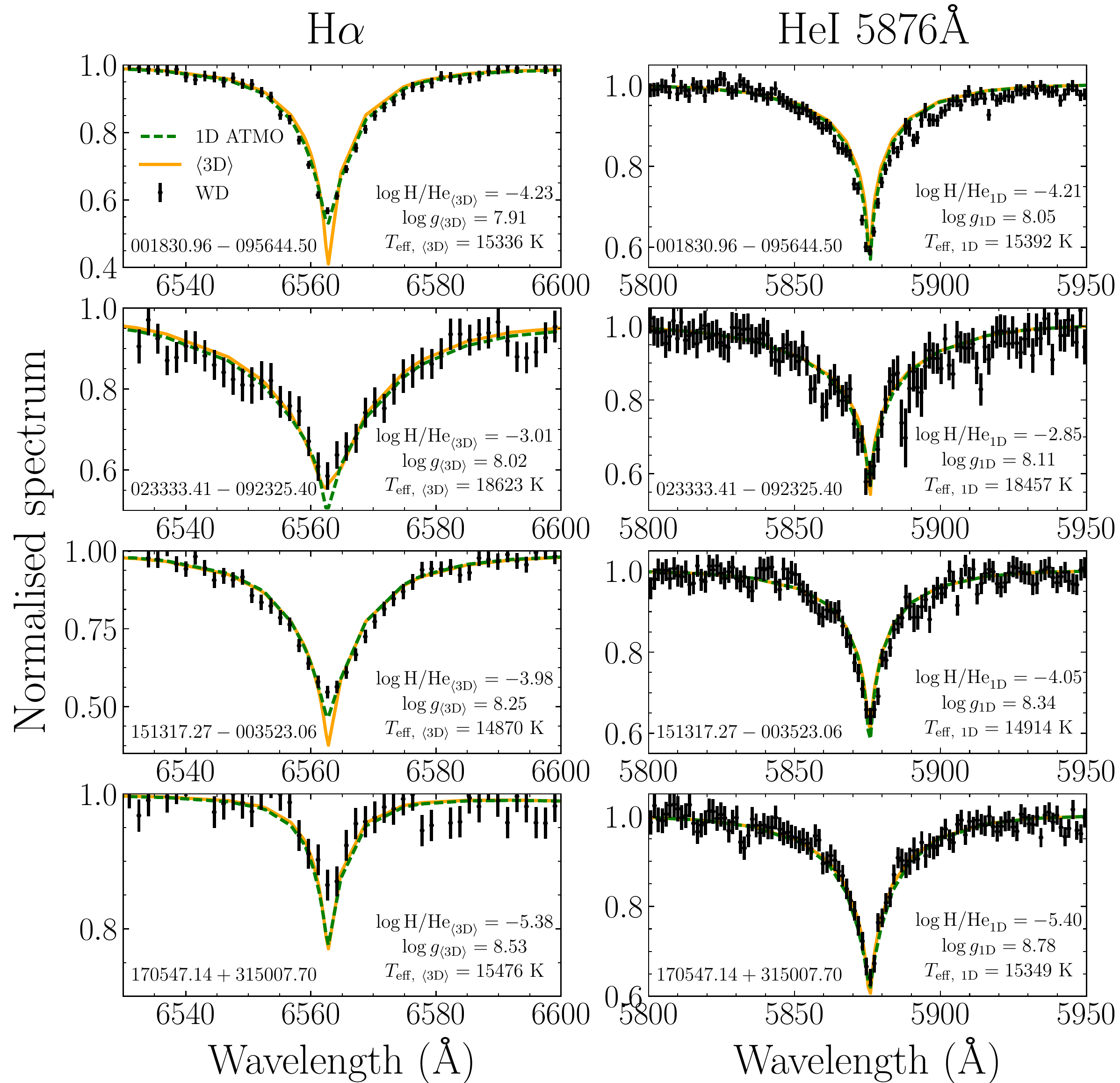}
    \caption{A comparison between several SDSS white dwarf spectra from \protect\cite{genest2019b} and 1D ATMO and \threeD\ synthetic spectra. The white dwarf spectra alongside their errors are shown in black. The white dwarfs were fitted with grids of \threeD\ and 1D spectra and the fitted atmospheric parameters are indicated on each sub plot. In solid orange and dashed green we plot the \threeD~and 1D fits. The sub plots on the left show the H$\alpha$ line, whereas the sub plots on the right are for the He~\textsc{I}~5876~\AA\ line. The two plots on each row are for the same white dwarf, with the name of the white dwarf indicated on each sub plot.} 
    \label{fig:obs_wd_cores}
\end{figure*}

To see how the synthetic and observed line cores compare, we selected a number of SDSS spectra from \cite{genest2019b}. These white dwarf spectra were fitted with grids of \threeD~and 1D spectra and the fitted parameters are indicated in the sub plots of Fig.~\ref{fig:obs_wd_cores}. The \threeD~and 1D fits to the observed spectra are also plotted on Fig~\ref{fig:obs_wd_cores}. The spectra are compared in the regions of H$\alpha$ and He~\textsc{I}~5876~\AA~lines. We chose this particular He~\textsc{I} line, because it is formed highest up in the atmosphere of all the He~\textsc{I} lines, and thus it is most likely to show issues with line cores.

We find that in some cases, the H$\alpha$ is deeper in 3D models when compared to observed spectra (the top and the two bottom sub plots on the left), but in three of the cases the 1D line core is also too deep (three bottom left sub plots). The He~\textsc{I}~5876~\AA~does not show significant disagreement in the line cores between the observations and the synthetic spectra.  The fact that H$\alpha$ shows issues with line cores, but He~\textsc{I}~5876~\AA line does not, agrees with the conclusions of \cite{tremblay_2013_spectra}, i.e. the deeper line cores are caused by overshoot in the upper layers, since the He~\textsc{I}~5876~\AA~is not formed as high up as the H$\alpha$ line.

Recently, \cite{klein2020} showed that DB white dwarfs in a particular effective temperature range show an emission core in the He~\textsc{I}~5876~\AA~line. They concluded that this cannot be explained by current models. This finding in combination with the information shown in Fig.~\ref{fig:obs_wd_cores} and in the results of \cite{tremblay_2013_spectra} for 3D DA white dwarfs, clearly indicates that there are some missing physics in both \threeD~and 1D models. However, it is clear that the \threeD~DB and DBA spectroscopic corrections are not affected by line core issues as demonstrated in Fig.\,\ref{fig:line_cores_removal} and \citet{myprecious}. Thus, we do not remove line cores in the rest of our analysis. Note also, Fig.~\ref{fig:obs_wd_cores} shows that although the 1D and \threeD~fits are of similar quality outside the line cores, the fitted atmospheric parameters, especially \logg, are significantly different.
 
\subsection{3D correction function}~\label{sec:threeD_corr}

For the ease of user application of 3D DB and DBA corrections, we provide correction functions that can be applied directly to spectroscopic \y, \logg~and \teff~parameters found using 1D synthetic spectra. This removes the need for users to interpolate the 3D DB and DBA corrections, as well as provides definitive corrections that do not vary between studies, since different studies can use non-identical interpolation methods. Our fitted corrections are not based in any physical arguments. Unlike interpolation, we do not aim to have a perfect fit between our correction functions and the 3D DB and DBA corrections. This is because we do not want to preserve the small fluctuations of our results, which could be the result of e.g. issues arising due to the finite size of the simulation or the effects of boundary conditions. 

To determine our correction functions we have written a code following the description of the recursive fitting procedure from \cite{ludwig1999}, \cite{sbordone2010} and \cite{allende2013}. This code not only provides the values of the fitted coefficients but also determines the function to fit. Our aim is to determine three correction functions for 3D \y, \logg~and \teff~corrections, in the form $f($\logg,~\teff,~\y$;\vec{A})$, where $\vec{A}$ is the vector of fitted coefficients. The code begins by fitting a simple function of $f($\logg,~\teff,~\y$;\vec{A}) = A_0$, where $A_0$ is the average of the corrections. The next step is then to replace $A_0$ with $(A_0 + A_1 \exp[A_2 + A_3 g_x + A_4 T_x + A_5 y_x])$, where
\begin{equation}~\label{eq:gxtxyx}
\begin{gathered}
g_x = (\log{g} - 7.0)/7.0,\\
T_x = (T_{\rm{eff}} - 10\,000.0)/10\,000.0,\\
y_x = -\log{\rm{(H/He)}}/(10.0).
\end{gathered}
\end{equation}
For this subsequent fit, the initial values are set as 
\begin{equation}~\label{eq:initial_coefficients}
\begin{gathered}
A_0 = A_0,\\
A_1 = 0.7 \times A_0,\\
A_2 = 0.5 \times A_0,\\
A_3 = 0.2 \times A_0,\\
A_4 = 0.1 \times A_0,\\
A_5 = 0.01 \times A_0,
\end{gathered}
\end{equation}
and are all based on the value of $A_0$ found during the first fit. At this point, we begin our recursive fitting procedure, where each coefficient of $A_i$ in the previous step is replaced one at a time by ($A_i + A_6 \exp[A_7 + A_8 g_x + A_9 T_x + A_{10} y_x]$), resulting in six separate minimisations. The initial values of the five new unknown coefficients are set as described in Eq.~\ref{eq:initial_coefficients}, but with $A_i$ replacing $A_0$. The best fitted correction function is then determined using the \texttt{least\_squares} module of \texttt{Python SciPy} package \citep{scipy2020}. This step is followed by a similar step where each parameter (11 at this point) is replaced by $A_j = A_j + A_{11} \exp[A_{12} + A_{13} g_x + A_{14} T_x + A_{15} y_x]$. The function with the smallest value of the cost function ($=0.5\chi^2$) is then chosen as the final fitted correction function. Note that in the case of 3D \logg~corrections, the fitting stopped at 11 parameters, as including more parameters resulted in over-fitting which was visually apparent. 

\begin{figure*}
	\includegraphics[width=1.5\columnwidth]{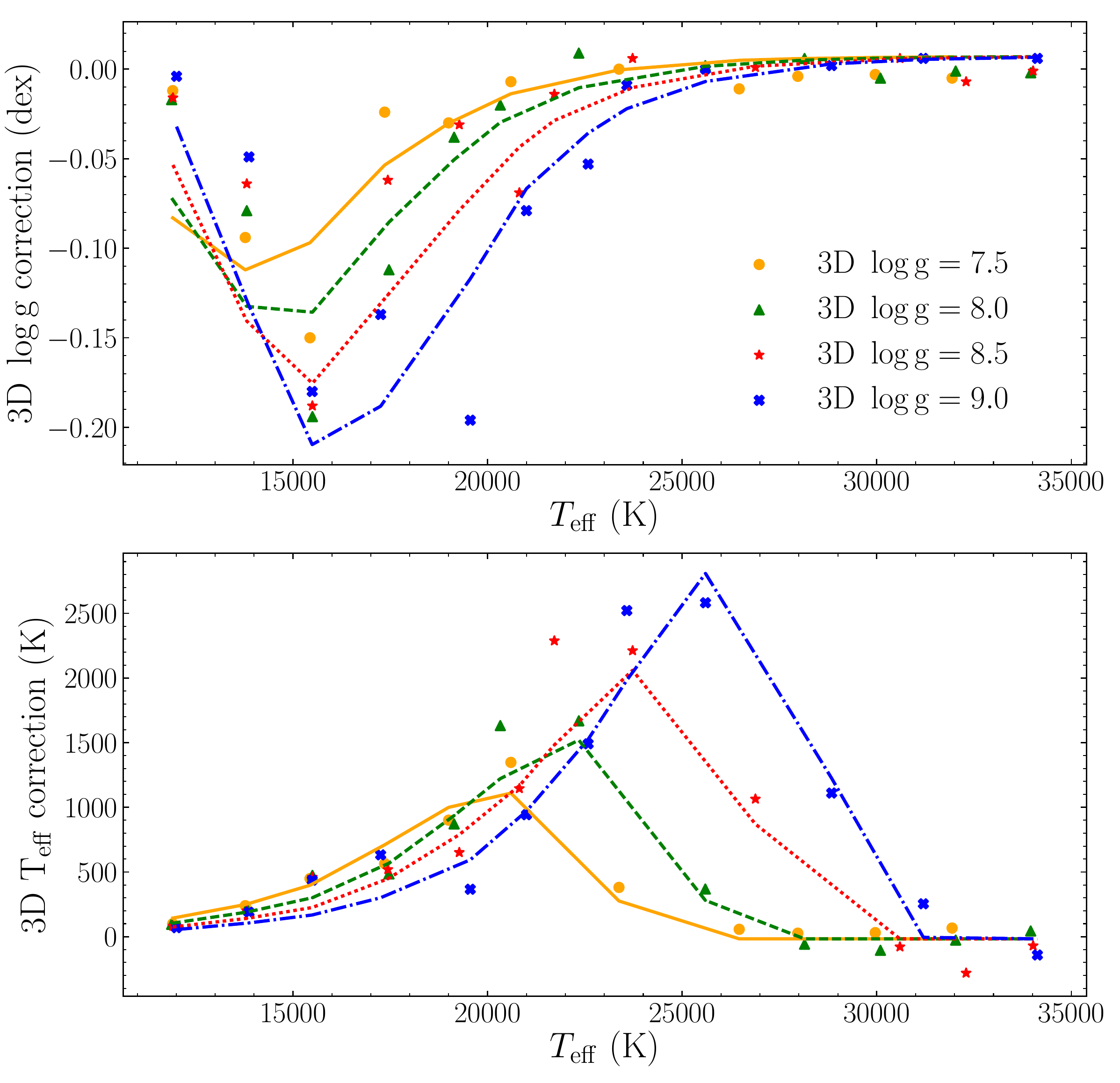}
    \caption{A comparison between the results of the correction functions and the 3D DBA corrections for the grid with \y~= $-$5. 3D corrections for \logg~$=7.5$, 8.0, 8.5 and 9.0 models are shown as orange circles, green triangles, red stars and blue diagonal crosses, respectively. The results of the correction functions for \logg~$=7.5$, 8.0, 8.5 and 9.0 models are shown as solid orange, dashed green, dotted red and dot-dashed blue lines, respectively. } 
    \label{fig:comp_corrs_with_corr_func}
\end{figure*}

In Fig.~\ref{fig:comp_corrs_with_corr_func} we compare the 3D DBA corrections with the predictions of the correction functions for \y~= $-5$. The correction functions are

\begin{equation}~\label{eq:logg_corr}
\begin{split}
\Delta{\log{g}} = a_0 + a_1 \exp \Big[a_2 + a_3 g_x + a_4 T_x + \Big(a_5 + a_6 \\ \times \exp \big[ a_7 + a_8 g_x + a_9 T_x + a_{10} y_x\big] \Big) y_x \Big],
\end{split}
\end{equation}

\begin{equation}~\label{eq:teff_corr}
\begin{split}
\Delta{T_{\rm{eff}}} = b_0 + b_1 \exp \Big[ \Big(b_2 + (b_6 + b_{11}  \\ \times \exp[b_{12} + b_{13} g_x + b_{14} T_x + b_{15} y_x]) \\ \times \exp\big[ b_7 + b_8 g_x + b_9 T_x + b_{10} y_x \big]\Big) + b_3 g_x + b_4 T_x + b_5 y_x \Big],
\end{split}
\end{equation}
where $\Delta{\log{g}}$ is the 3D \logg~correction and $\Delta{T_{\rm{eff}}}$ is the 3D \teff~correction. These corrections were derived using $g_x, T_x$ and $y_x$, therefore they have to be added in the following way to the 1D spectroscopically-determined parameters

\begin{equation}~\label{eq:how_to_add}
\begin{gathered}
\log{g}_{\rm{3D}} = \log{g}_{\rm{1D}} + 7 \times \Delta{\log{g}} \\
T_{\rm{eff, \ 3D}} = T_{\rm{eff, \ 1D}} + 10\,000 \times \Delta{T_{\rm{eff}}}.
\end{gathered}
\end{equation}

The 3D \y~corrections are insignificant, especially compared to typical observational errors and we do not discuss them further.
Tab.~\ref{tab:coeff_3d} gives the values of the fitted coefficients ($a_i$ and $b_i$). Note that outside the parameter range of our 3D corrections, these functions lose all meaning and should not be used. The parameter range for 1D derived spectroscopic values is $7.5 \leq$~\logg~$\leq 9.1$ dex, $11\,900 \leq$~\teff~$\leq 33\,900$ K and $-10.0 \leq$~\y~$\leq -2.0$ dex. In Appendix~\ref{sec:python_code} we supply a Python code to apply our correction functions. 

\begin{figure*}
	\includegraphics[width=2\columnwidth]{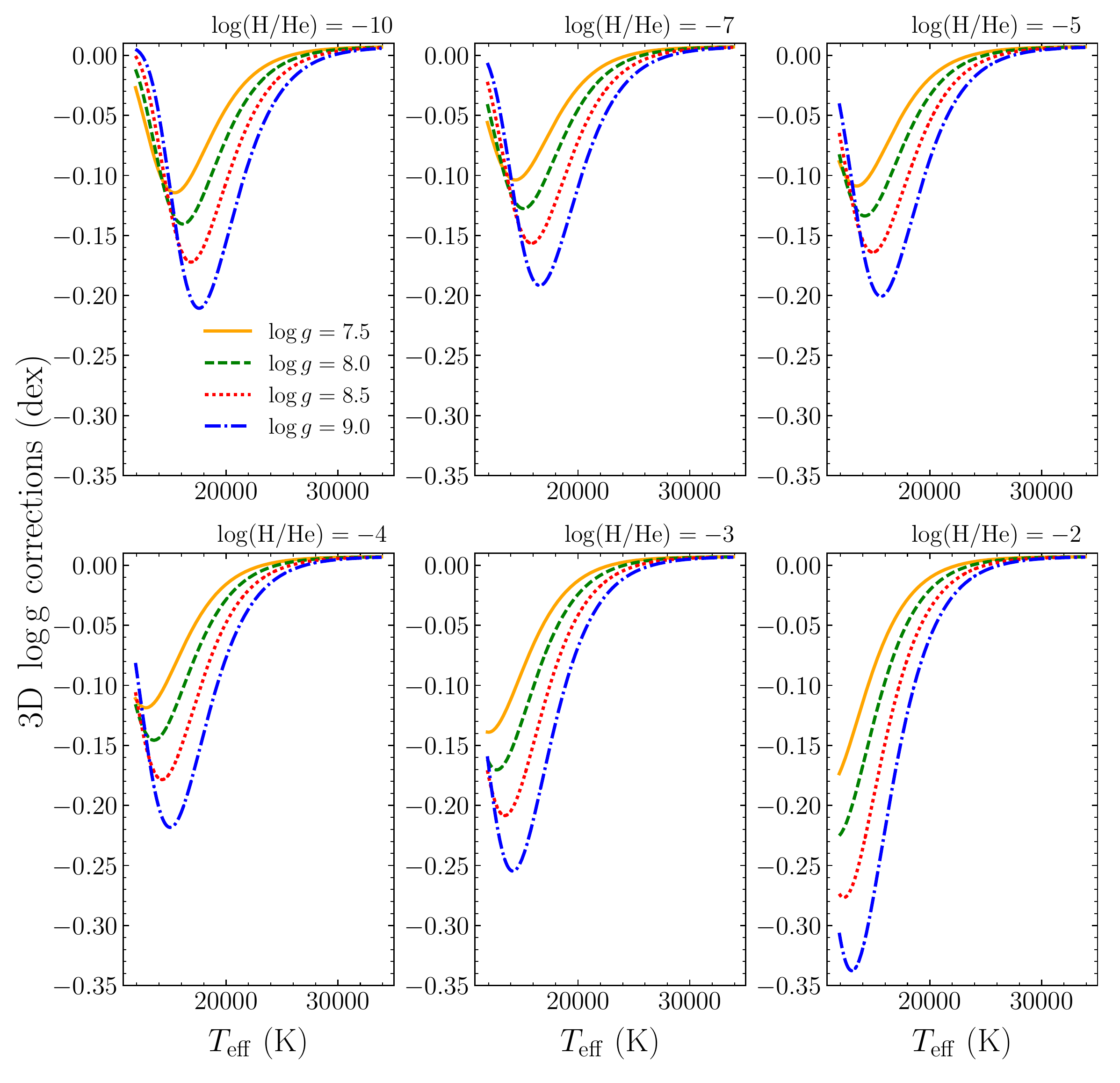}
    \caption{3D correction functions for \logg~shown for all \y, \logg~and \teff~values covered by our study. In solid orange, dashed green, dotted red and dot-dashed blue we show the \logg~corrections for \logg~$=7.5$, 8.0. 8.5 and 9.0 values, respectively. The abundances are indicated on each panel.} 
    \label{fig:3d_logg_corr}
\end{figure*}

\begin{figure*}
	\includegraphics[width=2\columnwidth]{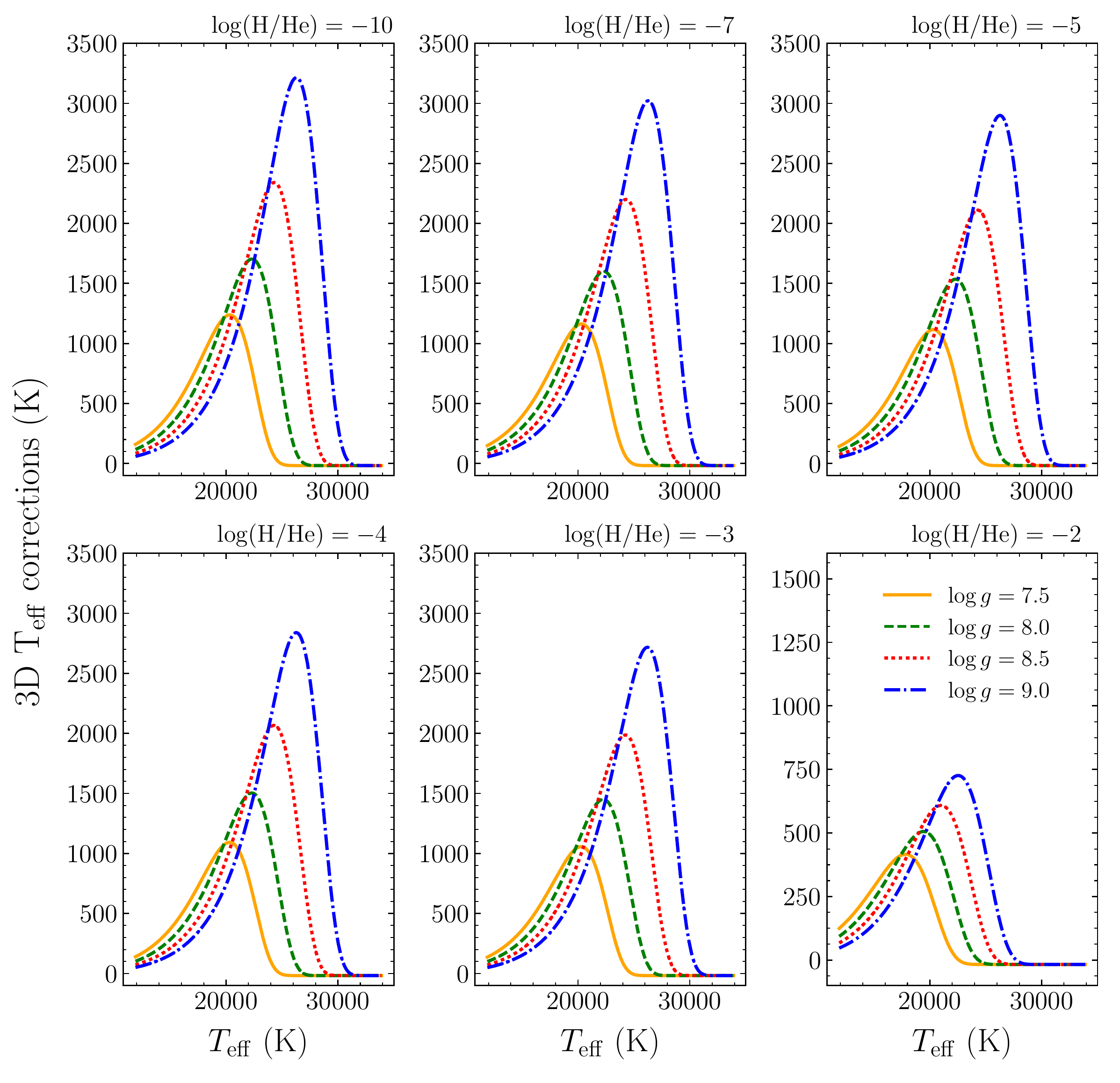}
    \caption{Same as Fig.~\ref{fig:3d_logg_corr}, but for 3D \teff~corrections.} 
    \label{fig:3d_teff_corr}
\end{figure*}

In Figs.~\ref{fig:3d_logg_corr} and~\ref{fig:3d_teff_corr} we show the 3D \logg~and \teff~correction functions for all values of \y, \logg~and \teff~considered in this study. There are significant 3D \logg~corrections for \teff~below around 20\,000~K depending on the hydrogen abundance, such that 3D synthetic spectra predict lower \logg~than 1D models. Uncertainties in the van der Waals broadening (discussed in the following section) fall in a similar parameter range \citep{beauchamp1996,bergeron_db_2011}, \teff~$ \leq 16\,000$~K, which overlaps well with our 3D \logg~corrections especially for \y~$\geq -4.0$ models. Significant \teff~corrections are observed for 18\,000~$\leq$~\teff~$\leq 28\,000$ K depending on the \logg~value. This is the temperature range where the issue of cool/hot solutions appears \citep{bergeron_db_2011}. In this region the He~\textsc{i} lines reach maximum strength, such that the they look identical with decreasing or increasing \teff~near this maximum point. This means that in this region the fitting becomes insensitive to \teff~and this could explain the significant 3D \teff~corrections. 
Nevertheless, the 3D \teff~corrections could have a strong effect on the spectroscopic parameters of the white dwarfs in the V777 Her (DBV) instability region. There is currently an issue with the empirical blue edge, which is too cool in comparison with observations by around 2\,000 K (at \logg~$\approx 8.0$) \citep{shipman2002,provencal2003,hermes2017,vangrootel2017}. However, our 3D \teff~corrections at \logg~$\approx 8.0$ and \teff~$\approx 31\,000$ K (the atmospheric parameters of the empirical blue edge) are insignificant and therefore cannot solve the disagreement between theory and observations. 

\begin{table}
	\centering
	\caption{The fitted coefficients of the 3D correction functions described in Eqs.~\ref{eq:logg_corr} and~\ref{eq:teff_corr}.}
	\label{tab:coeff_3d}
	\begin{tabular}{lrlr} 
		\hline
 		 Coeff. &   & Coeff. \\ 
		\hline
                $a_{0}$ & 1.032681e-03 & $b_{0}$ & $-$1.726331e-03 \\ 
                $a_{1}$ & $-$4.056833e-02 & $b_{1}$ & 2.018858e-02 \\ 
                $a_{2}$ & 2.224059e-01 & $b_{2}$ & $-$6.121790e-01 \\ 
                $a_{3}$ & 6.512899e+00 & $b_{3}$ & $-$3.942139e+00 \\ 
                $a_{4}$ & $-$3.736203e+00 & $b_{4}$ & 3.002973e+00 \\ 
                $a_{5}$ & 1.552502e+00 & $b_{5}$ & 1.974865e-01 \\ 
                $a_{6}$ & $-$2.384917e+00 & $b_{6}$ & $-$3.983912e+00 \\ 
                $a_{7}$ & 8.543144e-01 & $b_{7}$ & 5.171429e+00 \\ 
                $a_{8}$ & 3.556967e+00 & $b_{8}$ & $-$7.523787e+00 \\ 
                $a_{9}$ & $-$3.504215e+00 & $b_{9}$ & 3.786523e+00 \\ 
                $a_{10}$ & $-$1.751281e-02 & $b_{10}$ & $-$4.768727e+01 \\ 
                $a_{11}$ & - & $b_{11}$ & $-$3.889600e-04 \\ 
                $a_{12}$ & - & $b_{12}$ & $-$2.195071e+00 \\ 
                $a_{13}$ & - & $b_{13}$ & $-$6.955563e+00 \\ 
                $a_{14}$ & - & $b_{14}$ & 1.417272e+00 \\ 
                $a_{15}$ & - & $b_{15}$ & 4.767425e+01 \\ 
 		\hline
	\end{tabular}
\end{table}

\section{Discussion}~\label{observations}
For a demonstration of 3D corrections, the sample of SDSS DB and DBA white dwarfs is by far the largest spectroscopic sample available \citep{koester2015,kepler2019,gentilefusillo2019,genest2019b}. We rely specifically on the 1D spectroscopic parameters published in \cite{genest2019b}. We cross-matched this sample with the \textit{Gaia} DR2 white dwarf catalogue \citep{gentilefusillo2019,gentilefusillo2019cat}. As \cite{genest2019b} remarked, only around 90\% of DB and DBA white dwarfs can be cross-matched with the \textit{Gaia} white dwarf catalogue and we report a similar percentage. 
We removed all white dwarfs with spectroscopic signal-to-noise ratio (S/N) below 20. This results in a sample of 126 DB and 402 DBA white dwarfs. We require data with the highest precision possible to apply 3D corrections for an appropriate H/He ratio and in order to inspect systematic model issues.

\begin{figure*}
	\includegraphics[width=1.5\columnwidth]{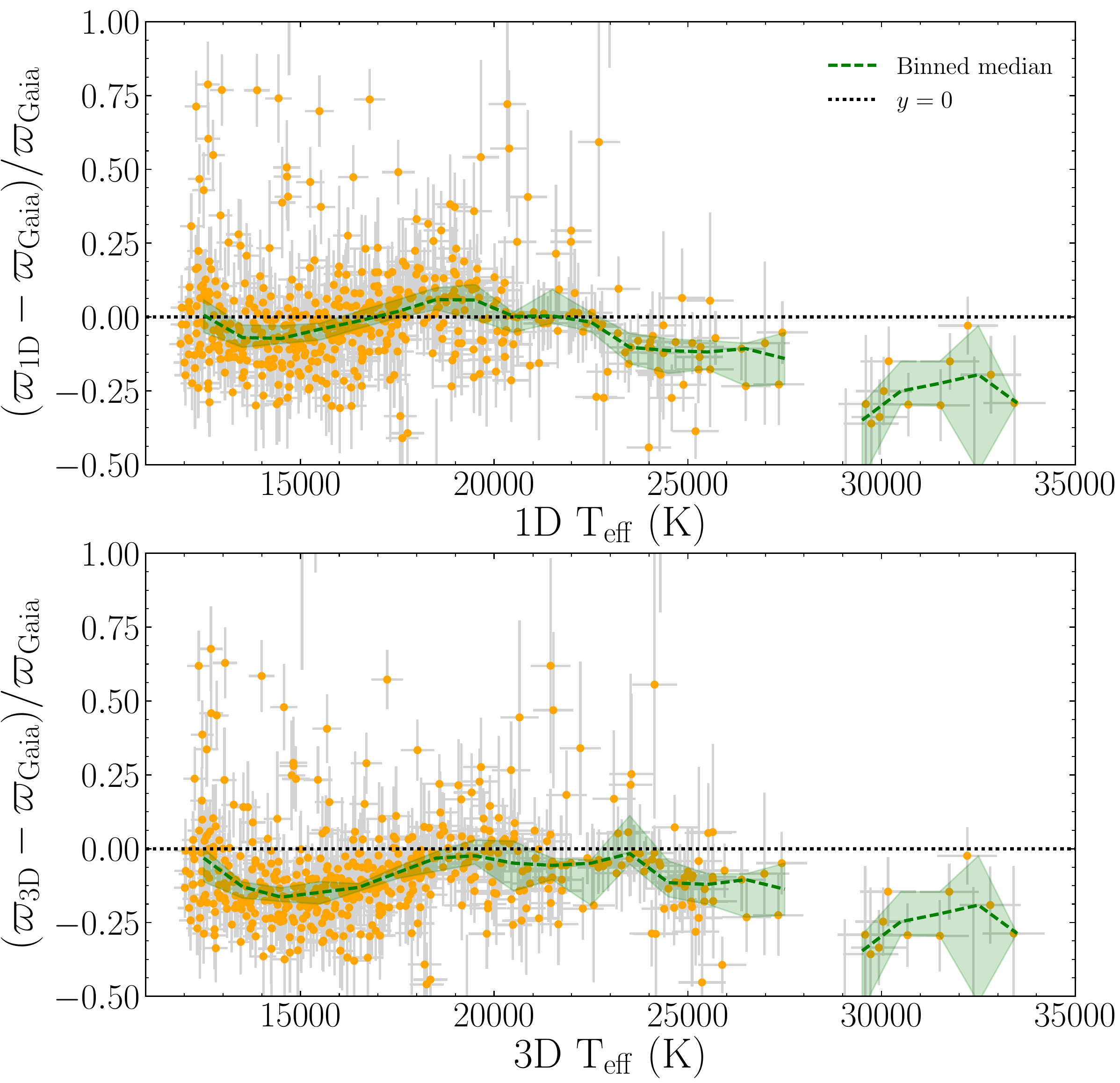}
    \caption{A comparison between the \textit{Gaia} parallaxes and the parallaxes derived from spectroscopic parameters without (top panel) and with (bottom panel) 3D DB and DBA corrections for the \protect\cite{genest2019b} SDSS sample. The orange filled circles represent fractional difference between the observed and theoretical parallax, $\varpi$, and in light grey we show the error on the difference. The dotted black line illustrates a perfect agreement. The running median of the fractional difference in bins of 1000 K is shown in dashed green. The block-coloured green area indicates the 95\% confidence limit of the median, which has been calculated using bootstrapping.}
    \label{fig:gaia_oneD_threeD_comp}
\end{figure*}

We have used 1D and 3D corrected spectroscopic parameters to predict synthetic \textit{Gaia} absolute $G$ magnitudes from our model atmospheres\footnote{We use 1D ATMO model atmospheres to predict absolute $G$ magnitudes in all cases as 3D effects on absolute fluxes are negligible.}. From the distance modulus linking the observed apparent \textit{Gaia} $G$ magnitude with our predicted absolute magnitude, we derive so-called 1D and 3D spectroscopic parallaxes. These values can be compared to observed parallaxes. Fig.~\ref{fig:gaia_oneD_threeD_comp} (upper panel) shows the comparison between 1D spectroscopic parallaxes and \textit{Gaia} parallaxes, which is similar to the results presented in \cite{genest2019b}. Overall, for individual white dwarfs the agreement is satisfactory within 1-3$\sigma$. In general, spurious large spectroscopic \logg~values should appear above zero on the figure because as \logg~increases at constant \teff, the absolute $G$ magnitude increases, i.e. the white dwarf becomes dimmer. Thus, for the same apparent $G$ magnitude the white dwarf must be closer, and therefore its spectroscopic parallax must be larger. If we take the median of the parallax difference in bins of 1\,000 K as shown in Fig.~\ref{fig:gaia_oneD_threeD_comp}, it is clear that the so-called high-\logg~problem is not apparent in the spectroscopic parallax distribution unlike the results of previous studies of DB and DBA white dwarfs such as \cite{koester2015} and \cite{rolland2018} as highlighted in \cite{tremblay2019}. In fact, it appears that the 1D spectroscopic results of \cite{genest2019b} may have some leaning towards low \logg~values. \citet{genest2019b} attribute the high-\logg~problem seen in earlier studies to the inclusion of DB and DBA white dwarfs with weak helium lines where the spectroscopic technique becomes unreliable, as well as the use of \cite{unsold_1955} treatment of van der Waals line broadening. Neither of these two issues affecting spectroscopic parameters are fully resolved and may also depend on the flux calibration and instrumental resolution of the observations.

Before we investigate this further, in Fig.~\ref{fig:gaia_oneD_threeD_comp} we also compare the 3D spectroscopic parallaxes with \textit{Gaia}. The 3D spectroscopic parallaxes where calculated as outlined previously, but with atmospheric parameters corrected for 3D effects. For individual white dwarfs, both 1D and 3D results are in satisfactory agreement with \textit{Gaia}, suggesting it is not possible to differentiate between 1D and 3D models on a case by case basis, although 3D parameters should be favoured as a starting point because of the superior input physics. When looking at the median in bins of 1\,000\,K, the bump which is seen in the 1D-\textit{Gaia} comparison at around 19\,000 K seems to largely disappear with the use of 3D models. The \teff~range where the bump is observed largely agrees with the \teff~range of 3D DB corrections discussed at length in \cite{myprecious} and shown in Fig.\,\ref{fig:3d_logg_corr}. At lower \teff~values, where the high-\logg~problem was historically reported, 3D models do not produce a better agreement with \textit{Gaia}, since the 1D \logg~values are already on average too low in comparison with \textit{Gaia} observations. 

\begin{figure*}
	\includegraphics[width=1.5\columnwidth]{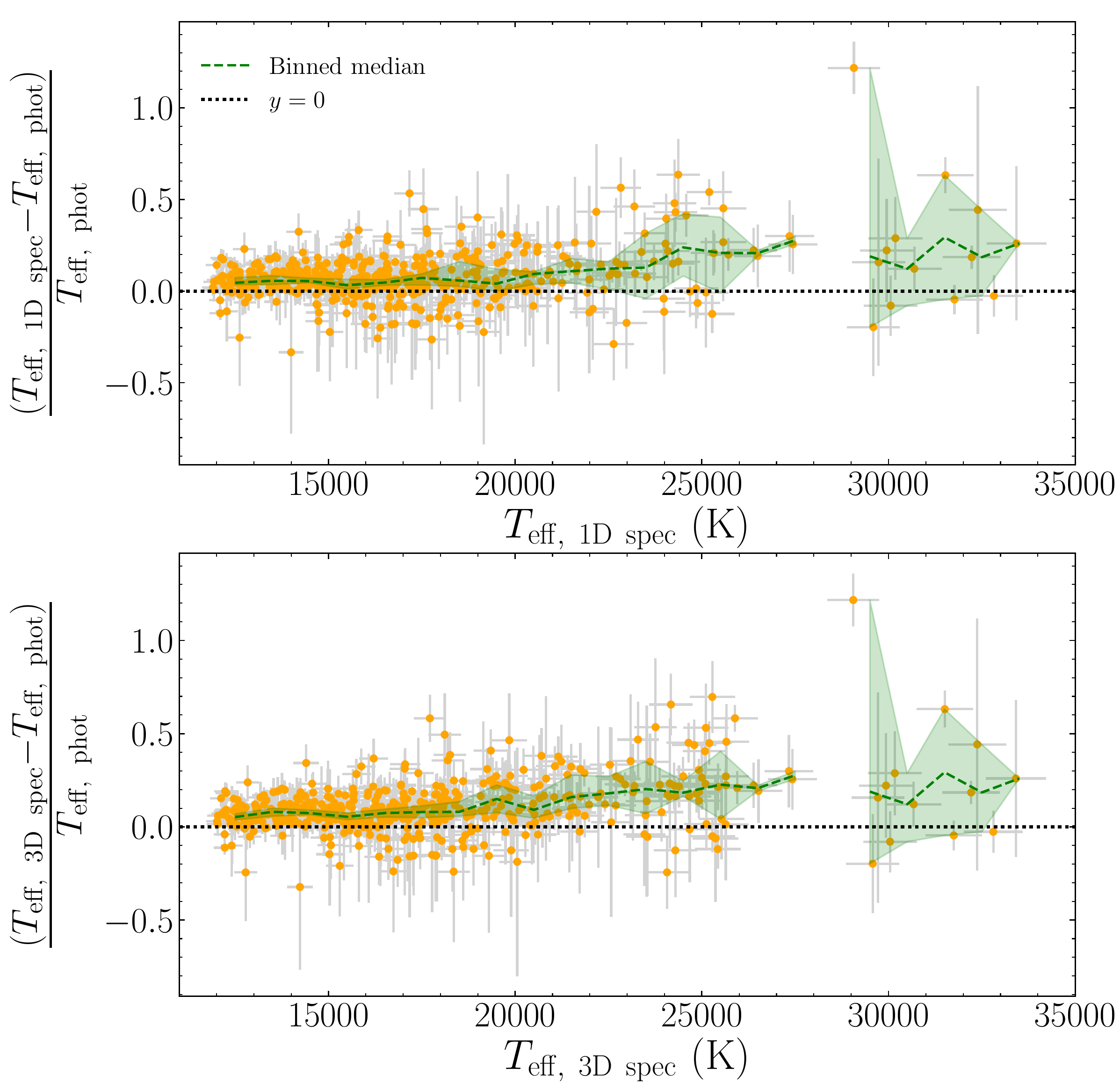}
    \caption{A comparison between the spectroscopically- and photometrically-determined values of \teff~for the \protect\cite{genest2019b} sample. The spectroscopic parameters have been computed without (top panel) and with (bottom panel) 3D \teff~corrections taken in to account. The photometric parameters are from \protect\cite{gentilefusillo2019}. They are calculated based on \textit{Gaia} data alone and include a reddening correction. The orange filled circles represent the fractional difference between the spectroscopic and photometric \teff, and the error on the difference is shown in light grey. The dotted black line illustrates a perfect agreement. The running median of the fractional difference in bins of 1000 K is shown in dashed green. The block-coloured green area indicates the 95\% confidence limit on the median.}
    \label{fig:cynthia_teff}
\end{figure*}

\begin{figure*}
	\includegraphics[width=1.5\columnwidth]{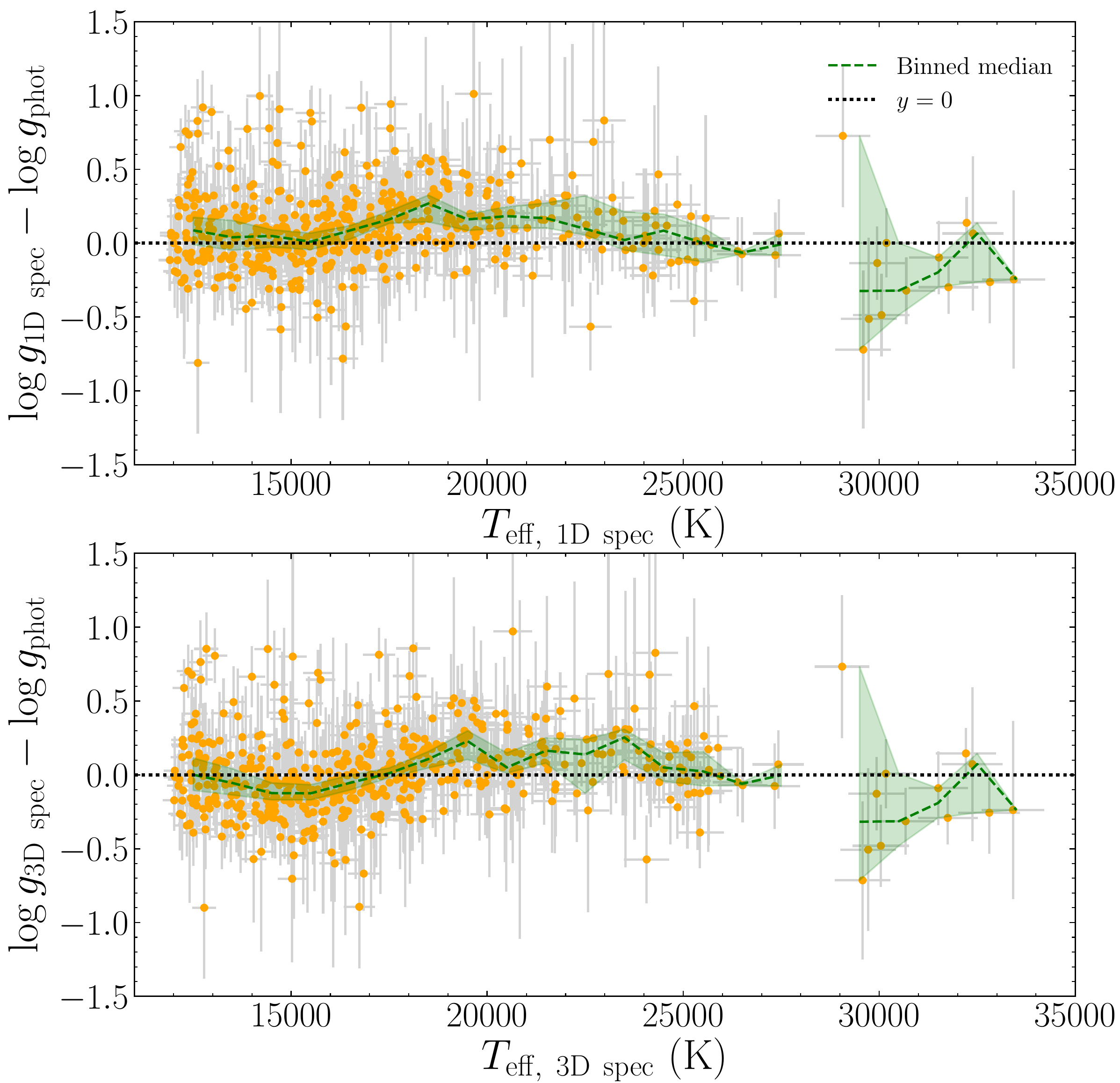}
    \caption{Similar to Fig.~\ref{fig:cynthia_teff} but for spectroscopically- and photometrically-determined values of \logg.}
    \label{fig:cynthia_logg}
\end{figure*}

The spectroscopic parallaxes were computed from a combination of \logg~and \teff~values as well as \textit{Gaia} $G$ magnitudes. In order to investigate the current status of the accuracy of spectroscopic parameters in better detail, in Figs.~\ref{fig:cynthia_teff} and~\ref{fig:cynthia_logg} we plot a comparison of \logg~and \teff~values determined from spectroscopic (with and without 3D corrections) and photometric \textit{Gaia} observations. 
The photometric parameters have been determined using only \textit{Gaia} data and have been extracted from the \textit{Gaia} DR2 white dwarf catalogue of \cite{gentilefusillo2019,gentilefusillo2019cat}. The photometric parameters are based on pure-helium models but the presence of hydrogen makes a negligible contribution to the \textit{Gaia} photometric parameters of DBA white dwarfs \citep{genest2019b}, which is not the case for cooler DC white dwarfs with much weaker helium opacity \citep{bergeron2019}. It was also previously shown that photometric parameters of DB and DBA white dwarfs have a fairly smooth $\log g$ distribution as a function of temperature, but the accuracy of the parameters is directly subject to the accuracy of photometric $G_{\rm BP}-G_{\rm RP}$ colour calibration \citep{tremblay2019}. This is an additional uncertainty that did not play a role in our comparison of spectroscopic parallaxes as it only depends on the absolute flux calibration which is tied to the observed flux of Vega \citep{bohlin2014}.

From Fig.~\ref{fig:cynthia_teff} it is apparent that both 1D and 3D spectroscopic models result in higher \teff~values when compared with photometrically derived values. It is unclear whether the offset is due to photometric colour calibration, SDSS spectral calibration, reddening or any issue with the spectroscopic parameters.
In Fig.~\ref{fig:cynthia_logg}, the comparison between \logg~values derived from spectroscopy and photometry is also shown. We confirm that within the errors of \cite{genest2019b} the high-\logg~problem previously observed in \citet{koester2015} and \citet{rolland2018} is non-existent. The effect of 3D corrections on $\log g$ values is fairly similar to that of 3D corrections on spectroscopic parallaxes observed in Fig.\,\ref{fig:gaia_oneD_threeD_comp}. This is not surprising given that the photometric $\log g$ distribution is fairly smooth as a function of temperature \citep{tremblay2019,genest2019a}, and fluctuations appear to be related to spectroscopic $\log g$ values, which are employed in both Figs.\,\ref{fig:gaia_oneD_threeD_comp} and \ref{fig:cynthia_logg}. 

We now attempt to understand better the systematic differences between photometric and spectroscopic studies by reviewing the input microphysics and comparing to external data from DA white dwarfs with vastly different microphysics. 

\subsection{van der Waals line broadening}~\label{sec:micro}

\begin{figure*}
	\includegraphics[width=1.5\columnwidth]{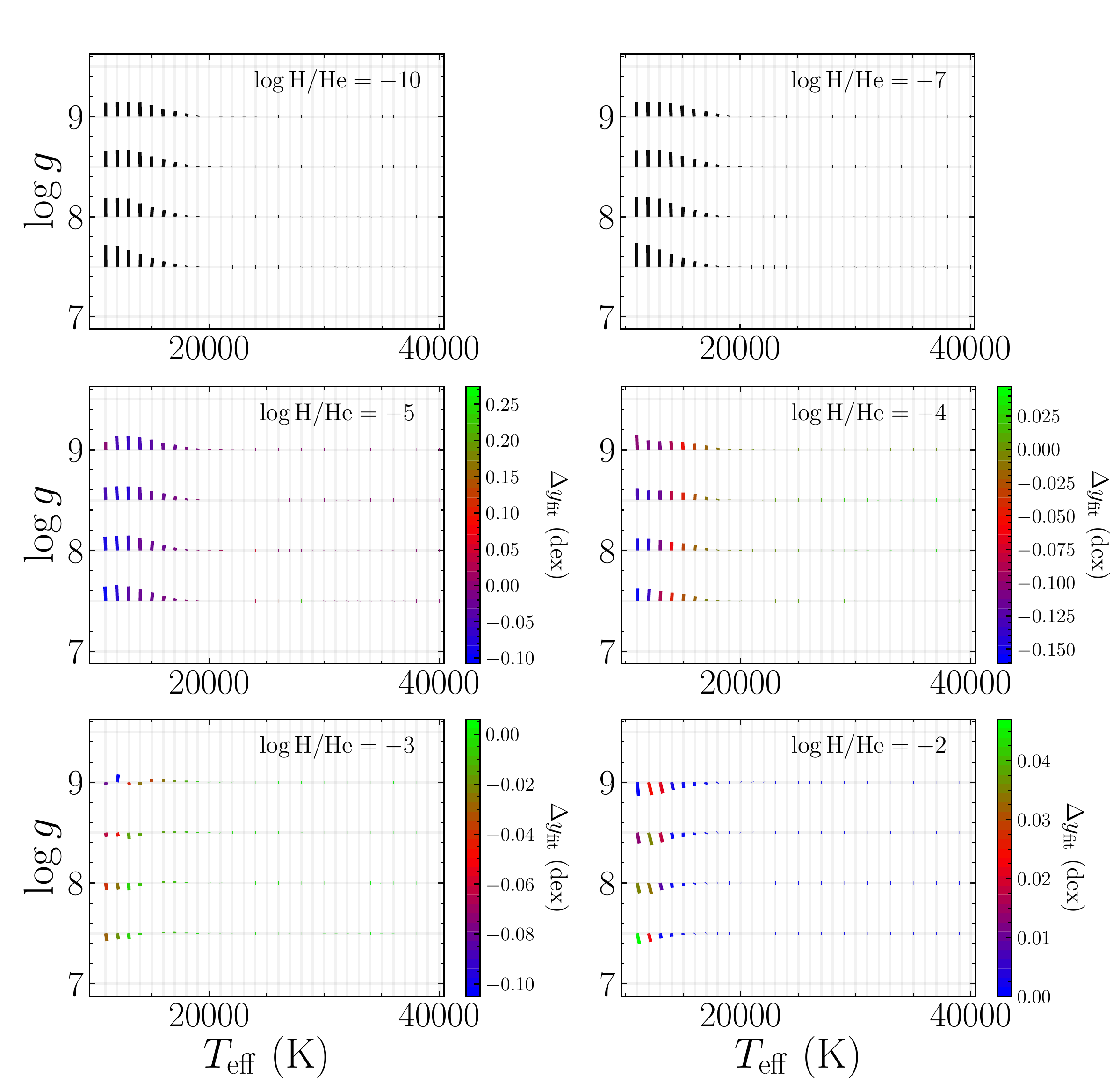}
    \caption{The corrections between two types of van der Waals line broadening. The intersections of the light grey lines denote the atmospheric parameters determined using the adapted \protect\cite{deridder1976} broadening. The coloured lines which extend from the intersections indicate the size of the correction. The end-point away from the intersection gives the values of the corresponding atmospheric parameters when \protect\cite{unsold_1955} broadening is used. The colours of the lines represent the \y~correction, which are omitted for very low hydrogen abundances.}
    \label{fig:vdw}
\end{figure*}

There are two types of van der Waals line broadening commonly used in spectroscopic analyses of DB and DBA white dwarfs. These are the \cite{unsold_1955} broadening, used in studies such as \cite{bergeron_db_2011}, and the modified line broadening of \cite{deridder1976} recently resurrected by \cite{genest2019a,genest2019b}. In this section, we aim to investigate the effect of van der Waals broadening on the values of the atmospheric parameters and to explain the disappearance of the high-\logg~problem. In order to achieve this we employ the 1D ATMO code to calculate two grids of models, one utilising the \cite{unsold_1955} broadening theory and the other using \cite{deridder1976} with the prescriptions of \cite{beauchamp1996}. We fit the latter grid with the former and in Fig.~\ref{fig:vdw} we show the van der Waals atmospheric parameter corrections. It is apparent that \logg~is most affected by the choice of the broadening, with \cite{unsold_1955} broadening resulting in larger values of \logg~for models with \y~$<-3.0$. This has already been noted by \cite{beauchamp1996}. 

To investigate further, we derive van der Waals correction functions for \y, \logg~and \teff~to transfrom from \cite{deridder1976} to \cite{unsold_1955} spectroscopically-determined parameters. We use the same technique as before. The resulting corrections are
\begin{equation}~\label{eq:logg_corr_vdw}
\begin{split}
\Delta{\log{g}}_{\rm{vdw}} = \Big( d_0 + d_6 \exp \Big[d_7 + d_8 g_x + d_9 T_x \\ + (d_{10} + d_{11} exp [d_{12} + d_{13} g_x + d_{14} T_x + d_{15} y_x]) y_x\Big]\Big)+ d_1 \exp [ \\ d_2 + d_3 g_x + d_4 T_x + d_5 y_x],
\end{split}
\end{equation}

\begin{equation}~\label{eq:teff_corr_vdw}
\begin{split}
\Delta{T_{\rm{eff, \ vdw}}} = \Big(e_0 + e_6 \exp \Big[e_7 + e_8 g_x + e_9 T_x \\ + (e_{10} + e_{11} \exp[e_{12} + e_{13} g_x + e_{14} T_x + e_{15} y_x]) y_x \Big]\Big) + e_1 \exp[ \\ e_2 + e_3 g_x + e_4 T_x + e_5 y_x] ,
\end{split}
\end{equation}
and the values of the fitted coefficients can be found in Table~\ref{tab:coeff_vdw}. As before, the corrections on \y~are negligible. 

\begin{table*}
	\centering
	\caption{The fitted coefficients of the van der Waals correction functions described in Eqs.~\ref{eq:logg_corr_vdw}~and~\ref{eq:teff_corr_vdw}.}
	\label{tab:coeff_vdw}
	\begin{tabular}{lrlr} 
		\hline
 		 Coeff. &   & Coeff. &   \\ 
		\hline
                $d_{0}$ & 5.844446e-05 & $e_{0}$ & 2.765480e-04 \\ 
                $d_{1}$ & $-$3.316185e+00 & $e_{1}$ & 5.078408e+02 \\ 
                $d_{2}$ & 4.709833e+00 & $e_{2}$ & 2.030607e+01 \\ 
                $d_{3}$ & 1.401743e+00 & $e_{3}$ & $-$2.762396e+00 \\ 
                $d_{4}$ & $-$5.285209e+00 & $e_{4}$ & $-$7.481534e+00 \\ 
                $d_{5}$ & $-$1.725072e+00 & $e_{5}$ & $-$1.417904e+02 \\ 
                $d_{6}$ & 3.097952e+00 & $e_{6}$ & $-$4.104236e-02 \\ 
                $d_{7}$ & 4.778176e+00 & $e_{7}$ & 6.809893e+00 \\ 
                $d_{8}$ & 1.401012e+00 & $e_{8}$ & 9.994226e+00 \\ 
                $d_{9}$ & $-$5.283561e+00 & $e_{9}$ & $-$1.903287e+01 \\ 
                $d_{10}$ & $-$5.097423e+00 & $e_{10}$ & 1.247415e+01 \\ 
                $d_{11}$ & 1.362931e+00 & $e_{11}$ & $-$8.114644e+00 \\ 
                $d_{12}$ & 9.057032e-01 & $e_{12}$ & 1.691382e+00 \\ 
                $d_{13}$ & 7.534572e-04 & $e_{13}$ & 1.460991e+00 \\ 
                $d_{14}$ & $-$1.824838e-03 & $e_{14}$ & $-$2.023648e+00 \\ 
                $d_{15}$ & 7.971757e-05 & $e_{15}$ & $-$3.645338e-01 \\ 
 		\hline
	\end{tabular}
\end{table*}

\begin{figure*}
	\includegraphics[width=1.5\columnwidth]{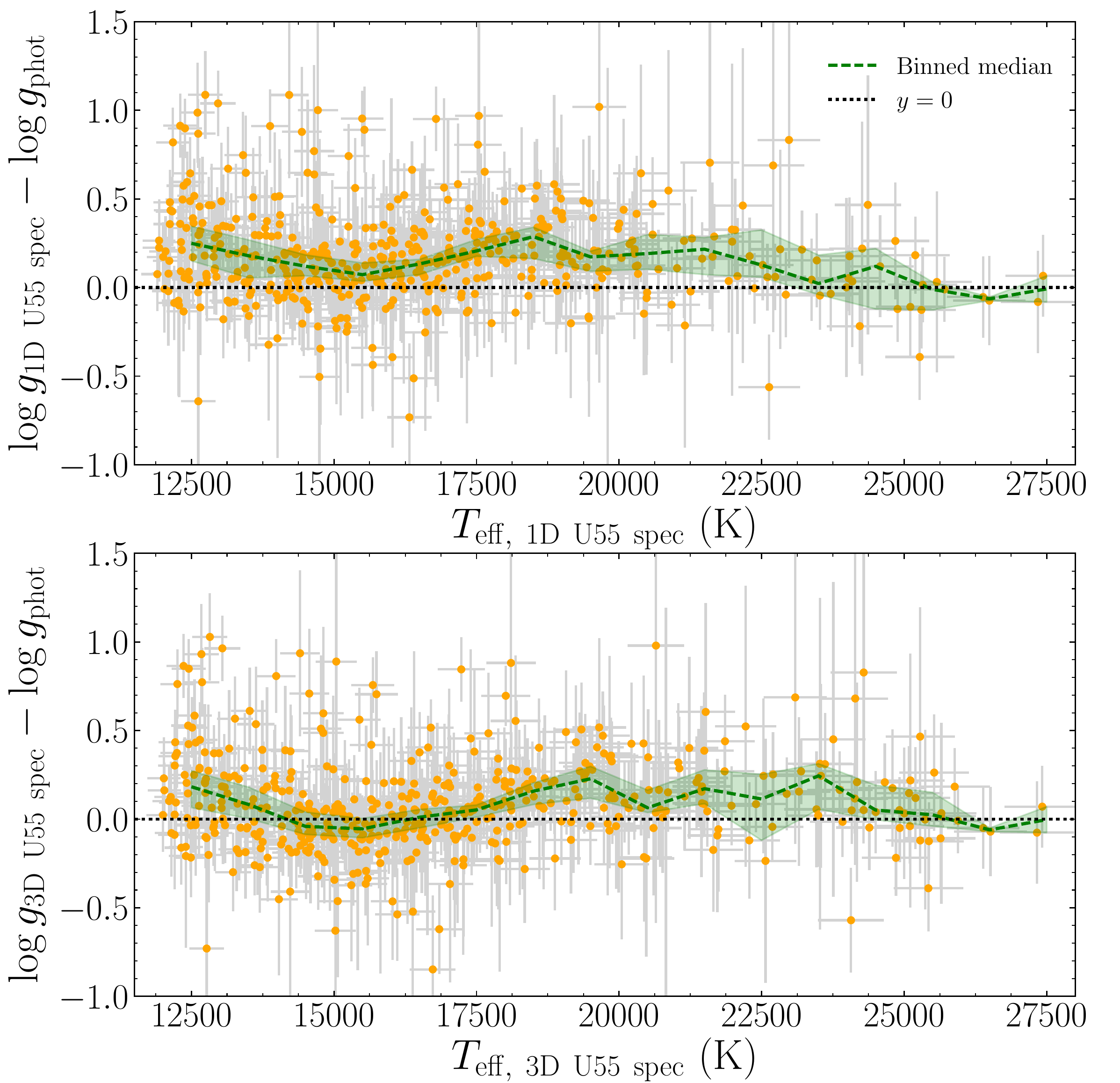}
    \caption{A comparison between the spectroscopically and photometrically-derived \logg~values corrected for van der Waals broadening only (top plot) and corrected both for van der Waals broadening and 3D effects (bottom plot). Solid orange circles represent the difference in \logg~with errors shown in light grey. The running median in bins of 1000 K is shown in dashed green, with the 95\% confidence limit being represented by the green colour blocked area. For reference, the dotted black line illustrates a one-to-one agreement.}
    \label{fig:cynthia_vdw}
\end{figure*}

As an illustration we apply these corrections to the \cite{genest2019b} sample. Recall that \cite{genest2019b} sample uses the \cite{deridder1976} broadening. Due to the way the corrections were derived, they have to be subtracted from the values of the atmospheric parameters derived using the \cite{deridder1976} broadening theory, such that

\begin{equation}~\label{eq:how_to_add2}
\begin{gathered}
\log{g}_{\rm{U}} = \log{g}_{\rm{D}} - 7 \times \Delta{\log{g}}_{\rm{vdw}} \\
T_{\rm{eff, \ U}} = T_{\rm{eff, \ D}} - 1000 \times \Delta{T_{\rm{eff, \ vdw}}},
\end{gathered}
\end{equation}
where D and U denote the parameters derived using models with \cite{deridder1976} and \cite{unsold_1955} line broadenings, respectively.
In Fig.~\ref{fig:cynthia_vdw} we show a comparison between the photometric and van der Waals corrected spectroscopic \logg~values. From Figs.~\ref{fig:vdw} and~\ref{fig:cynthia_vdw}, it is clear that the \cite{unsold_1955} theory does lead to higher values of \logg~at low \teff. 
Using \cite{deridder1976} broadening is, however, not the final answer, since this theory was adapted in the DB and DBA case to better fit observations when using 1D model atmospheres \citep{beauchamp1996}. Instead, the aim should be to determine a better prescription of van der Waals line broadening. Additionally, it is therefore not surprising that 1D models produce a slightly better agreement with \textit{Gaia} observations compared to 3D models when using \cite{deridder1976}. In Fig.~\ref{fig:cynthia_vdw} we also show a comparison between spectroscopic and photometric \logg~values, when the SDSS spectroscopic parameters of \cite{genest2019b} are corrected for both the van der Waals broadening and 3D effects. We see that in this case, the 3D models show a slightly better agreement with observations, hinting that \cite{unsold_1955} van der Waals broadening is closer to the real prescription of the broadening. However, this combination of corrections still leaves an irregular $\log g$ distribution below \teff~$\approx 14\,000$ K. We conclude that 3D corrections should be employed alongside properly adjusted line broadening using Eqs.~\ref{eq:logg_corr_vdw}~and~\ref{eq:teff_corr_vdw} until a better prescription is developed.

\subsection{Non-ideal effects}

\begin{figure*}
	\includegraphics[width=1.5\columnwidth]{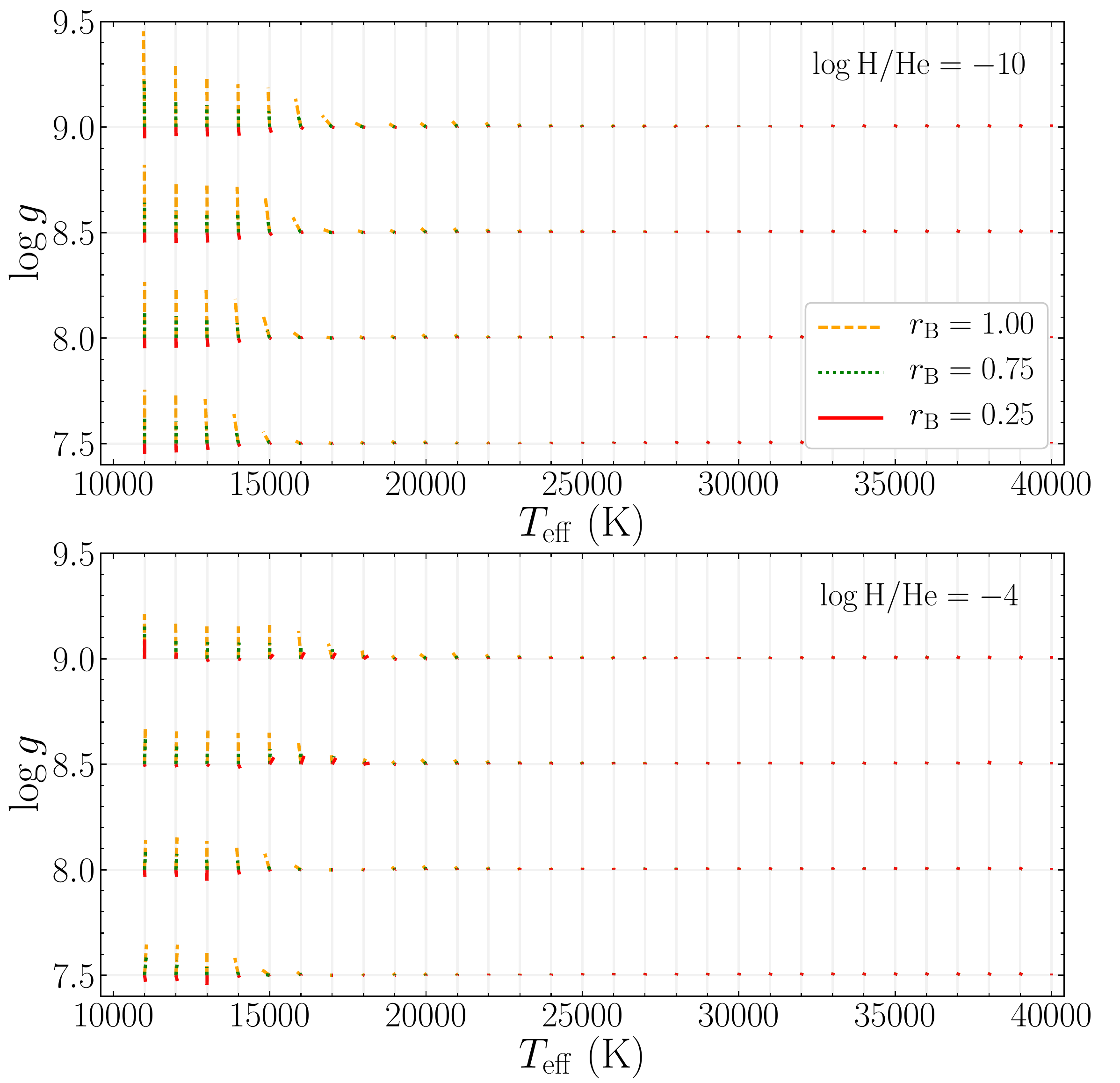}
    \caption{The corrections in \logg~and \teff~arising from varying the value of the multiplying factor to the Bohr radius, $r_{\rm{B}}$, in the \citet{hummer1988} non-ideal gas theory with respect to the standard value of 0.5. In solid orange, dotted green and dashed red we show the corrections for $r_{\rm{B}} = 1.0$, 0.75 and 0.25, respectively. The \y~value is indicated on each panel.}
    \label{fig:non_ideal}
\end{figure*}

In this section we derive corrections due to non-ideal gas perturbations from neutral atoms (i.e. neutral helium) on the atomic levels of light-absorbing helium atoms. To do this, we calculate new grids of 1D ATMO models with different parameterisations of the \citet{hummer1988} theory currently used in all DB model atmospheres. We use different multiplicative factors to the Bohr radius \rb, namely \rb=[0.25,0.75,1.00], which are then used to scale the size of neutral helium atom. The standard 1D ATMO grid used in previous sections was calculated at \rb~$=0.5$ and we shall use this grid as a reference for fitting. All previous studies using the ATMO code have relied on this parameterisation \citep{bergeron_db_2011,rolland2018,myprecious,genest2019a,genest2019b}. The derived corrections for \logg~and \teff~for select DB and DBA grids are shown in Fig.~\ref{fig:non_ideal}. We omit the \y~corrections because they are insignificant, at most a few per cent. Similarly to van der Waals broadening corrections, we find a significant effect on \logg~values. Increasing the value of \rb~from 0.5 to 1.0 results in an increase of around 0.4 dex in most extreme cases. For the highest hydrogen abundances  (\y~$=-2.0$), we find that the non-ideal effects do not change with varying value of \rb. The reason why \rb~does not seem to have much effect on this particular \y~grid is because the hydrogen lines are so strong that they overwhelm the fitting in that particular range of \teff. 

Because the agreement of current 1D and 3D spectroscopic parameters with \textit{Gaia} is reasonable and the non-ideal and line broadening corrections are partially degenerate, we argue that the commonly used value of \rb~$=0.5$ is still an optimal choice. However, a more physical treatment of non-ideal effects will be needed before we can verify the accuracy of 3D corrections in this regime.

\subsection{Comparison between He- and H-atmosphere white dwarfs}

A comparison of the atmospheric parameter distributions for both DA and DB/DBA white dwarfs can help to understand systematic trends. Line broadening physics is dramatically different between the two spectral types, i.e. in warmer DA white dwarfs hydrogen is broadened by the linear Stark effect, while helium is subject to the quadratic Stark effect and van der Waals broadening. As a consequence, if there were any issues caused by microphysics we expect systematic trends to be different between spectroscopic parameters of DA and DB stars. In contrast \textit{Gaia} photometric parameters have a much weaker dependence on atmospheric composition above 12\,000\,K \citep{bergeron2019} and the median masses of DA and DB stars are found to be the same to within a few per cent \citep{tremblay2019}.

\begin{figure*}
	\includegraphics[width=1.5\columnwidth]{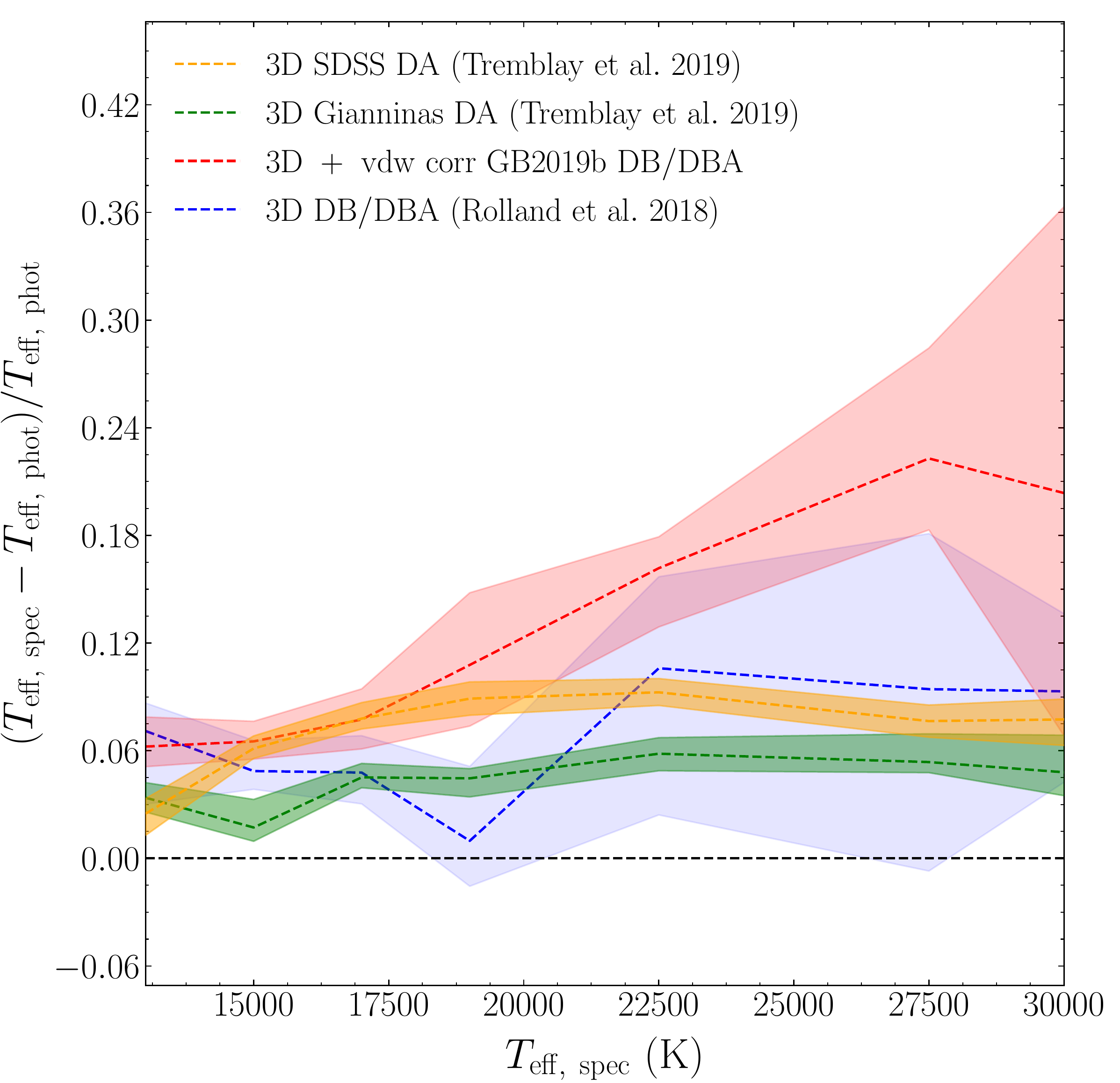}
    \caption{A comparison between the \teff~derived using spectroscopic and \textit{Gaia} photometric observations for samples of DA and DB/DBA white dwarfs. The median fractional difference of each sample was plotted in bins of 2000\,K for $T_{\rm eff} \leq $ 20\,000\,K and of 5000 K above that temperature. The difference in \teff~for the SDSS DA sample of \protect\cite{tremblay2019}, with spectroscopic S/N $>$ 20 and 3D corrections, is shown in dashed orange. The dashed green curve corresponds to the DA sample of \protect\cite{gianninas2011} with 3D corrections (see \citealt{tremblay2019} for the comparison with \textit{Gaia}). In dashed red and dashed blue the difference is shown for the SDSS DB/DBA sample of \protect\cite{genest2019b} and the DB/DBA sample of \protect\cite{rolland2018}, respectively, both corrected for 3D effects presented in this work. The \protect\cite{genest2019b} sample is also corrected for \protect\citet{unsold_1955} van der Waals broadening (using corrections from Section~\ref{sec:micro}), while \protect\cite{rolland2018} is already using this type of line broadening. The coloured areas represent the corresponding 95\% confidence limit on the medians calculated using bootstrapping. The dashed black line indicates a perfect agreement between spectroscopy and photometry.}
    \label{fig:da_teff}
\end{figure*}

In Fig.~\ref{fig:da_teff} we compare the \teff~derived from spectroscopic and \textit{Gaia} photometric observations of several different samples of DA and DB/DBA white dwarfs. In the plot we show the binned median of the fractional difference of each sample in bins of 2\,000 K for $12\,000 \leq $~\teff~$ \leq 20\,000$ K, and bins of 5\,000 K for \teff~$ > 20\,000$~K.
We rely on the same DB/DBA sample of \citet{genest2019b} as discussed previously, using our 3D corrections and also the \citet{unsold_1955} prescription of van der Waals broadening (with original atmospheric parameters corrected according to Section~\ref{sec:micro}). We also use the DB/DBA sample of \cite{rolland2018} where we have applied our new 3D DB/DBA corrections. The sample of \cite{rolland2018} is already using the \citet{unsold_1955} prescription of van der Waals broadening. Our DA white dwarf samples are drawn from \citet{gianninas2011} and SDSS \citep{tremblay2019}. For SDSS sample we imposed the restriction to spectroscopic S/N > 20, the same as for the DB and DBA samples. In both cases we have applied 3D DA corrections \citep{tremblay_2013_spectra}. These DA samples are effectively the same as the 3D spectroscopic samples described in \citet{tremblay2019}. All spectroscopic samples have been cross matched with the \textit{Gaia} white dwarf catalogue of \cite{gentilefusillo2019,gentilefusillo2019cat} to obtain photometric atmospheric parameters based on dereddened photometry.

The different samples show similar offsets between photometric and spectroscopic \teff. The offset is not obviously caused by calibration issues of SDSS spectra \citep{kleinman04,tremblay2019}, since the samples of \cite{gianninas2011} and \cite{rolland2018} do not use SDSS data. An issue with the approximate treatment of dereddening in \citet{gentilefusillo2019} is unlikely because there is no obvious correlation between the observed offset and distance \citep{tremblay2019}, e.g. the offset is similar even for bright DA white dwarfs within 40\,pc \citep{tremblay2020} for which reddening is expected to be negligible. It cannot be due to 3D spectroscopic effects, as 1D DB/DBA models predict a similar offset (see Fig.~\ref{fig:cynthia_teff}) and 3D effects for DA white dwarfs are essentially negligible above a temperature of 13\,000\,K. It is unlikely to be caused by microphysics issues, such as van der Waals or Stark broadening, as the offset seems to be more or less constant over the entire \teff~range and is very similar for DA and DB stars, whereas line opacities vary significantly as a function of \teff\ and spectral type. Therefore, they are unlikely to cause offsets of similar magnitude.


Therefore, this leaves the possibility that the offset is due to calibration issues with \textit{Gaia} colours, which are the direct input in the determination of photometric \teff\ values, given that the sensitivity of surface gravity to colours is weak. Similar offsets have been observed in other studies of photometric \textit{Gaia} data \citep{jesus2018,tremblay2019,genest2019a,tremblay2020} but our work provides more robust constraints based on 3D spectroscopic parameters of both DA and DB white dwarfs. The calibration of \textit{Gaia} astrometry \citep{lindergren2018} is not expected to have a significant role in the determination of surface temperatures. Finally, the larger scatter observed for DB/DBA samples above \teff~$\approx 20\,000$ K could be explained by spectral fitting issues regarding the maximum strength of He \textsc{i} lines as discussed in this work. For DA white dwarfs spectral fitting is straightforward in this temperature range. 

\begin{figure*}
	\includegraphics[width=1.5\columnwidth]{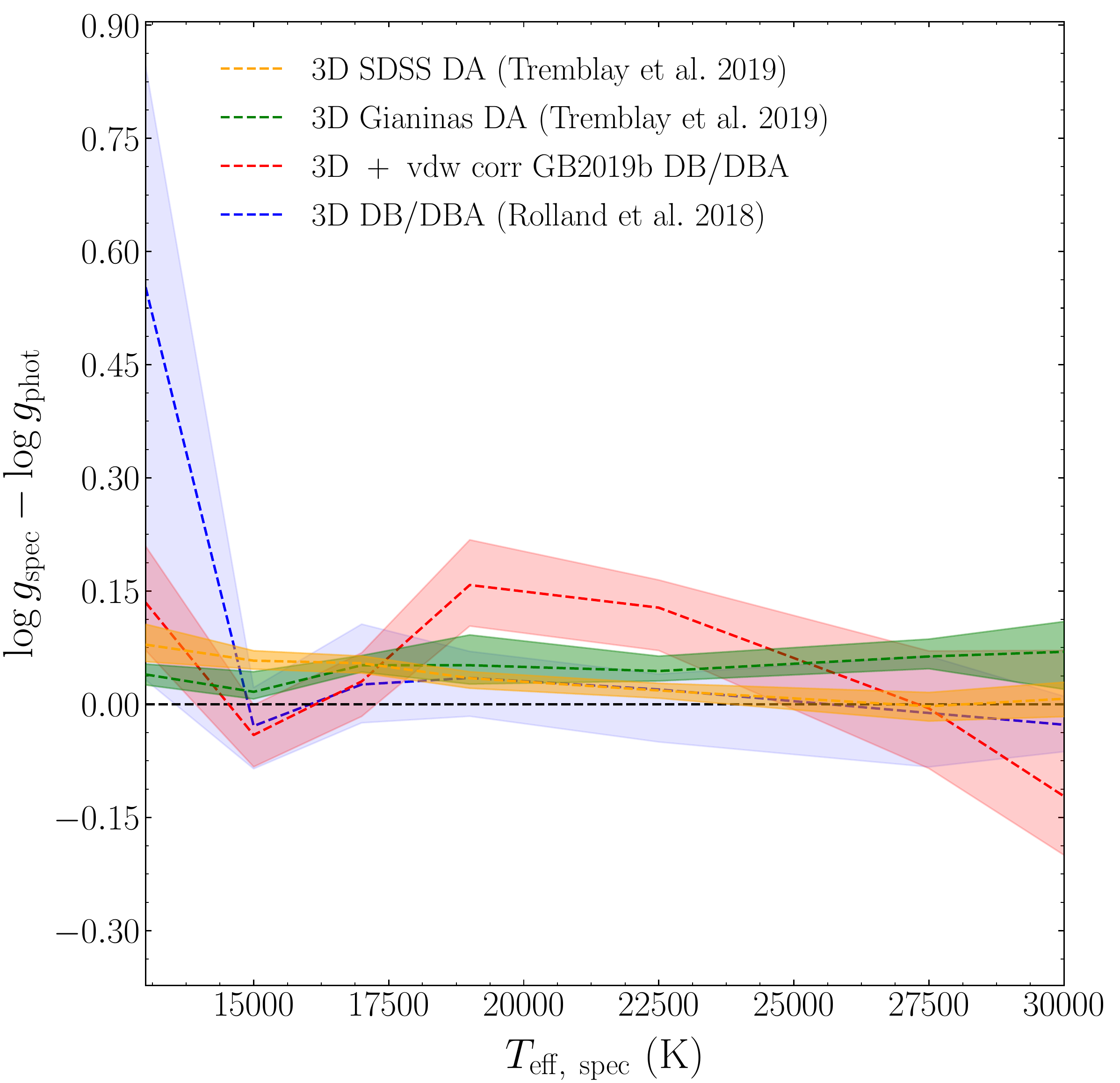}
    \caption{Same as Fig.~\ref{fig:da_teff} but for a comparison between the \logg~derived using spectroscopic and \textit{Gaia} photometric observations for samples of DA and DB/DBA white dwarfs.}
    \label{fig:da_logg}
\end{figure*}

In Fig.~\ref{fig:da_logg} we show a comparison between the spectroscopically- and photometrically-derived \logg~values. The remnant high-\logg~issue for DB and DBA white dwarfs can be seen in the (corrected for 3D effects and van der Waals broadening according to \citealt{unsold_1955}) \cite{rolland2018} and \cite{genest2019b} samples below \teff~$\approx 15\,000$~K. Similarly to the \teff~comparison, the agreement between spectroscopically- and photometrically-derived \logg~is not perfect. When deriving the photometric atmospheric parameters, the temperature almost only depends on observed colours, while for a fixed temperature value and mass-radius relation, the surface gravity only depends on mean absolute flux.
Therefore, it means that if an offset is observed in \teff~and is caused by \textit{Gaia} colour calibration, then an offset similar in shape is likely to be seen in \logg~values, as the radius must compensate for the offset in temperature to match absolute fluxes. However, the diagnostic potential is complicated by the fact that spectroscpic \teff\ and \logg\ values have different sensitivities and possibly different systematics with respect to the line profiles. For DA white dwarfs in Fig.~\ref{fig:da_logg} we find that photometric $\log g$ values are systematically lower. This results in larger radii, which compensate for the lower photometric $T_{\rm eff}$ values in keeping the same absolute fluxes, and is therefore entirely consistent with a colour calibration issue. In contrast the DB white dwarfs in both samples show an irregular behaviour, which could suggest that issues with line profiles dominate or are similar in strength to colour calibration issues.

In general, it appears that when comparing the spectroscopic results to external constraints, both DA and DB/DBA white dwarfs behave in a similar fashion. This indicates that 3D DB/DBA atmospheric models are comparable to their DA counterparts in terms of precision.

\section{Conclusions}~\label{sec:conclusions}

Using 282 3D atmospheric models of DB and DBA white dwarfs, we have determined the corrections for the spectroscopically-derived atmospheric parameters of \y, \logg~and \teff. These corrections are due to a more physical treatment of convection in 3D models when compared with their 1D counterparts. We find significant \logg~corrections in the \teff~range where the high-\logg~problem was historically reported for these types of white dwarfs. When applying our 3D corrections to the spectroscopic sample of \cite{genest2019b} we find a similar agreement between 1D and 3D spectroscopic parameters when compared with \textit{Gaia} data. We nevertheless recommend using 3D parameters as a standard starting point because of the superior input physics. We provide 3D correction functions that are differential and can be applied to 1D atmospheric parameters from any study and with any input model atmospheres. We provide an example \textsc{Python} code for application of the correction functions. 

The currently employed \cite{deridder1976} theory of van der Waals broadening has been specifically adapted to produce a smooth distribution of 1D atmospheric parameters for \teff~$\lesssim$ 16,000\,K, albeit this has not been updated for the most recent constraints from \textit{Gaia} DR2. 
Nevertheless, it is not surprising 
that 3D corrections do not lead to a better agreement with \textit{Gaia} photometry and astrometry in this regime. When applying 3D corrections to spectroscopically derived values relying on the \cite{unsold_1955} theory of van der Waals broadening, we find that 3D results are in better agreement with \textit{Gaia}.
However, we stress that the treatment of non-ideal effects due to neutral helium atoms also plays a significant role in this low temperature regime and that it is degenerate with the choice of the line broadening theory. 
This highlights the fact that the treatment of the microphysics for cool DB and DBA white dwarfs needs to be revisited. For example, after making a major improvement in the treatment of Stark broadening of neutral helium atoms using computer simulations, \cite{patricktremblay2020} also plan to extend their simulations to improve van der Waals broadening. 

By comparing spectroscopic and photometric atmospheric parameters of various samples of DA, DB and DBA white dwarfs, we have been able to identify a prominent offset in \teff~and a possible smaller offset in \logg. By ruling out a number of possibilities that could be responsible for such an offset, we conclude that it is most likely caused by the \textit{Gaia} colour calibration. A similar offset has been reported in other studies. In general, it seems that the offsets are remarkably similar for both DA and DB/DBA white dwarfs. Thus, the atmospheric models of DB and DBA white dwarfs can be considered to be of similar precision and accuracy to that of DA models. Additionally, the same test can employed in the future to assess the colour calibration of Gaia DR3.

\section*{Acknowledgements}
This project has received funding from the European Research Council (ERC) under the European Union's Horizon 2020 research and innovation programme (grant agreement No 677706 - WD3D). B.F. has been supported by the Swedish Research Council (Vetenskapsr{\aa}det). H.G.L. acknowledges financial support by the Sonderforschungsbereich SFB\,881 ``The Milky Way System'' (subprojects A4) of the German Research Foundation (DFG).

\section{Data Availability}
The observational data used this article are published in \cite{gianninas2011,rolland2018,gentilefusillo2019,gentilefusillo2019cat,tremblay2019,genest2019a} and \cite{genest2019b}. 
Simulation data underlying this article will be shared on reasonable request to the corresponding author.
The final 3D correction functions derived in this work are available in the article and in its online supplementary material.




\bibliographystyle{mnras}
\bibliography{aamnem99,aabib} 

\begin{thebibliography}{}
\makeatletter
\relax
\def\mn@urlcharsother{\let\do\@makeother \do\$\do\&\do\#\do\^\do\_\do\%\do\~}
\def\mn@doi{\begingroup\mn@urlcharsother \@ifnextchar [ {\mn@doi@}
  {\mn@doi@[]}}
\def\mn@doi@[#1]#2{\def\@tempa{#1}\ifx\@tempa\@empty \href
  {http://dx.doi.org/#2} {doi:#2}\else \href {http://dx.doi.org/#2} {#1}\fi
  \endgroup}
\def\mn@eprint#1#2{\mn@eprint@#1:#2::\@nil}
\def\mn@eprint@arXiv#1{\href {http://arxiv.org/abs/#1} {{\tt arXiv:#1}}}
\def\mn@eprint@dblp#1{\href {http://dblp.uni-trier.de/rec/bibtex/#1.xml}
  {dblp:#1}}
\def\mn@eprint@#1:#2:#3:#4\@nil{\def\@tempa {#1}\def\@tempb {#2}\def\@tempc
  {#3}\ifx \@tempc \@empty \let \@tempc \@tempb \let \@tempb \@tempa \fi \ifx
  \@tempb \@empty \def\@tempb {arXiv}\fi \@ifundefined
  {mn@eprint@\@tempb}{\@tempb:\@tempc}{\expandafter \expandafter \csname
  mn@eprint@\@tempb\endcsname \expandafter{\@tempc}}}

\bibitem[\protect\citeauthoryear{{Allende Prieto}, {Koesterke}, {Ludwig},
  {Freytag}  \& {Caffau}}{{Allende Prieto} et~al.}{2013}]{allende2013}
{Allende Prieto} C.,  {Koesterke} L.,  {Ludwig} H.~G.,  {Freytag} B.,
  {Caffau} E.,  2013, \mn@doi [\aap] {10.1051/0004-6361/201220064}, \href
  {https://ui.adsabs.harvard.edu/abs/2013A&A...550A.103A} {550, A103}

\bibitem[\protect\citeauthoryear{{Althaus}, {C{\'o}rsico}, {Isern}  \&
  {Garc{\'{\i}}a-Berro}}{{Althaus} et~al.}{2010}]{althaus2010}
{Althaus} L.~G.,  {C{\'o}rsico} A.~H.,  {Isern} J.,   {Garc{\'{\i}}a-Berro} E.,
   2010, \mn@doi [\aapr] {10.1007/s00159-010-0033-1}, \href
  {http://adsabs.harvard.edu/abs/2010A%26ARv..18..471A} {18, 471}

\bibitem[\protect\citeauthoryear{{Beauchamp}, {Wesemael}, {Bergeron}, {Liebert}
   \& {Saffer}}{{Beauchamp} et~al.}{1996}]{beauchamp1996}
{Beauchamp} A.,  {Wesemael} F.,  {Bergeron} P.,  {Liebert} J.,   {Saffer}
  R.~A.,  1996, in {Jeffery} C.~S.,  {Heber} U.,  eds,  Astronomical Society of
  the Pacific Conference Series Vol. 96, Hydrogen deficient stars.. p.~295

\bibitem[\protect\citeauthoryear{{Beauchamp}, {Wesemael}, {Bergeron},
  {Fontaine}, {Saffer}, {Liebert}  \& {Brassard}}{{Beauchamp}
  et~al.}{1999}]{beauchamp_1999_v777_dba}
{Beauchamp} A.,  {Wesemael} F.,  {Bergeron} P.,  {Fontaine} G.,  {Saffer}
  R.~A.,  {Liebert} J.,   {Brassard} P.,  1999, \mn@doi [\apj]
  {10.1086/307148}, \href {http://adsabs.harvard.edu/abs/1999ApJ...516..887B}
  {516, 887}

\bibitem[\protect\citeauthoryear{{B{\'e}dard}, {Bergeron}, {Brassard}  \&
  {Fontaine}}{{B{\'e}dard} et~al.}{2020}]{bedard2020}
{B{\'e}dard} A.,  {Bergeron} P.,  {Brassard} P.,   {Fontaine} G.,  2020,
  \mn@doi [\apj] {10.3847/1538-4357/abafbe}, \href
  {https://ui.adsabs.harvard.edu/abs/2020ApJ...901...93B} {901, 93}

\bibitem[\protect\citeauthoryear{{Bergeron}, {Wesemael}, {Michaud}  \&
  {Fontaine}}{{Bergeron} et~al.}{1988}]{bergeron1988}
{Bergeron} P.,  {Wesemael} F.,  {Michaud} G.,   {Fontaine} G.,  1988, \mn@doi
  [\apj] {10.1086/166705}, \href
  {https://ui.adsabs.harvard.edu/abs/1988ApJ...332..964B} {332, 964}

\bibitem[\protect\citeauthoryear{{Bergeron}, {Wesemael}  \&
  {Fontaine}}{{Bergeron} et~al.}{1991}]{bergeron1991}
{Bergeron} P.,  {Wesemael} F.,   {Fontaine} G.,  1991, \mn@doi [\apj]
  {10.1086/169624}, \href
  {https://ui.adsabs.harvard.edu/abs/1991ApJ...367..253B} {367, 253}

\bibitem[\protect\citeauthoryear{{Bergeron} et~al.,}{{Bergeron}
  et~al.}{2011}]{bergeron_db_2011}
{Bergeron} P.,  et~al., 2011, \mn@doi [ApJ] {10.1088/0004-637X/737/1/28}, \href
  {2011ApJ...737...28B} {737, 28}

\bibitem[\protect\citeauthoryear{{Bergeron}, {Dufour}, {Fontaine}, {Coutu},
  {Blouin}, {Genest-Beaulieu}, {B{\'e}dard}  \& {Rolland }}{{Bergeron}
  et~al.}{2019}]{bergeron2019}
{Bergeron} P.,  {Dufour} P.,  {Fontaine} G.,  {Coutu} S.,  {Blouin} S.,
  {Genest-Beaulieu} C.,  {B{\'e}dard} A.,   {Rolland } B.,  2019, \mn@doi
  [\apj] {10.3847/1538-4357/ab153a}, \href
  {https://ui.adsabs.harvard.edu/abs/2019ApJ...876...67B} {876, 67}

\bibitem[\protect\citeauthoryear{{Blouin}, {Dufour}, {Thibeault}  \&
  {Allard}}{{Blouin} et~al.}{2019}]{blouin2019}
{Blouin} S.,  {Dufour} P.,  {Thibeault} C.,   {Allard} N.~F.,  2019, \mn@doi
  [\apj] {10.3847/1538-4357/ab1f82}, \href
  {https://ui.adsabs.harvard.edu/abs/2019ApJ...878...63B} {878, 63}

\bibitem[\protect\citeauthoryear{{Bohlin}, {Gordon}  \& {Tremblay}}{{Bohlin}
  et~al.}{2014}]{bohlin2014}
{Bohlin} R.~C.,  {Gordon} K.~D.,   {Tremblay} P.~E.,  2014, \mn@doi [\pasp]
  {10.1086/677655}, \href
  {https://ui.adsabs.harvard.edu/abs/2014PASP..126..711B} {126, 711}

\bibitem[\protect\citeauthoryear{{B{\"o}hm-Vitense}}{{B{\"o}hm-Vitense}}{1958}]{bohm1958}
{B{\"o}hm-Vitense} E.,  1958, \zap, \href
  {http://adsabs.harvard.edu/abs/1958ZA.....46..108B} {46, 108}

\bibitem[\protect\citeauthoryear{{Caffau} \& {Ludwig}}{{Caffau} \&
  {Ludwig}}{2007}]{caffau_2007_lhd}
{Caffau} E.,  {Ludwig} H.-G.,  2007, \mn@doi [\aap]
  {10.1051/0004-6361:20077234}, \href
  {http://adsabs.harvard.edu/abs/2007A%26A...467L..11C} {467, L11}

\bibitem[\protect\citeauthoryear{{Cukanovaite}, {Tremblay}, {Freytag}, {Ludwig}
   \& {Bergeron}}{{Cukanovaite} et~al.}{2018}]{myprecious}
{Cukanovaite} E.,  {Tremblay} P.-E.,  {Freytag} B.,  {Ludwig} H.-G.,
  {Bergeron} P.,  2018, \mn@doi [\mnras] {10.1093/mnras/sty2383}, \href
  {http://adsabs.harvard.edu/abs/2018MNRAS.481.1522C} {481, 1522}

\bibitem[\protect\citeauthoryear{{Cukanovaite}, {Tremblay}, {Freytag},
  {Ludwig}, {Fontaine}, {Brassard}, {Toloza}  \& {Koester}}{{Cukanovaite}
  et~al.}{2019}]{mysunandstars}
{Cukanovaite} E.,  {Tremblay} P.~E.,  {Freytag} B.,  {Ludwig} H.~G.,
  {Fontaine} G.,  {Brassard} P.,  {Toloza} O.,   {Koester} D.,  2019, \mn@doi
  [\mnras] {10.1093/mnras/stz2656}, \href
  {https://ui.adsabs.harvard.edu/abs/2019MNRAS.490.1010C} {490, 1010}

\bibitem[\protect\citeauthoryear{{Cunningham}, {Tremblay}, {Gentile Fusillo},
  {Hollands}  \& {Cukanovaite}}{{Cunningham} et~al.}{2020}]{cunningham2020}
{Cunningham} T.,  {Tremblay} P.-E.,  {Gentile Fusillo} N.~P.,  {Hollands} M.,
  {Cukanovaite} E.,  2020, \mn@doi [\mnras] {10.1093/mnras/stz3638}, \href
  {https://ui.adsabs.harvard.edu/abs/2020MNRAS.492.3540C} {492, 3540}

\bibitem[\protect\citeauthoryear{{Deridder} \& {van Renspergen}}{{Deridder} \&
  {van Renspergen}}{1976}]{deridder1976}
{Deridder} G.,  {van Renspergen} W.,  1976, \aaps, \href
  {https://ui.adsabs.harvard.edu/abs/1976A&AS...23..147D} {23, 147}

\bibitem[\protect\citeauthoryear{{Doyle}, {Young}, {Klein}, {Zuckerman}  \&
  {Schlichting}}{{Doyle} et~al.}{2019}]{doyle2019}
{Doyle} A.~E.,  {Young} E.~D.,  {Klein} B.,  {Zuckerman} B.,   {Schlichting}
  H.~E.,  2019, \mn@doi [Science] {10.1126/science.aax3901}, \href
  {https://ui.adsabs.harvard.edu/abs/2019Sci...366..356D} {366, 356}

\bibitem[\protect\citeauthoryear{{Eisenstein} et~al.,}{{Eisenstein}
  et~al.}{2006}]{eisenstein2006}
{Eisenstein} D.~J.,  et~al., 2006, \mn@doi [\aj] {10.1086/504424}, \href
  {https://ui.adsabs.harvard.edu/abs/2006AJ....132..676E} {132, 676}

\bibitem[\protect\citeauthoryear{{Fantin} et~al.,}{{Fantin}
  et~al.}{2019}]{fantin2019}
{Fantin} N.~J.,  et~al., 2019, \mn@doi [\apj] {10.3847/1538-4357/ab5521}, \href
  {https://ui.adsabs.harvard.edu/abs/2019ApJ...887..148F} {887, 148}

\bibitem[\protect\citeauthoryear{{Fontaine} \& {Wesemael}}{{Fontaine} \&
  {Wesemael}}{1987}]{fontaine1987}
{Fontaine} G.,  {Wesemael} F.,  1987, in {Philip} A.~G.~D.,  {Hayes} D.~S.,
  {Liebert} J.~W.,  eds, IAU Colloq. 95: Second Conference on Faint Blue Stars.
  pp 319--326

\bibitem[\protect\citeauthoryear{{Fouesneau}, {Rix}, {von Hippel}, {Hogg}  \&
  {Tian}}{{Fouesneau} et~al.}{2019}]{fouesneau2019}
{Fouesneau} M.,  {Rix} H.-W.,  {von Hippel} T.,  {Hogg} D.~W.,   {Tian} H.,
  2019, \mn@doi [\apj] {10.3847/1538-4357/aaee74}, \href
  {https://ui.adsabs.harvard.edu/abs/2019ApJ...870....9F} {870, 9}

\bibitem[\protect\citeauthoryear{{Freytag}}{{Freytag}}{2013}]{freytag2013}
{Freytag} B.,  2013, Memorie della Societa Astronomica Italiana Supplementi,
  \href {http://adsabs.harvard.edu/abs/2013MSAIS..24...26F} {24, 26}

\bibitem[\protect\citeauthoryear{{Freytag}}{{Freytag}}{2017}]{freytag2017}
{Freytag} B.,  2017, \memsai, \href
  {http://adsabs.harvard.edu/abs/2017MmSAI..88...12F} {88, 12}

\bibitem[\protect\citeauthoryear{Freytag, Steffen, Ludwig, Wedemeyer-B{\"o}hm,
  Schaffenberger  \& Steiner}{Freytag et~al.}{2012}]{freytag2012_cobold}
Freytag B.,  Steffen M.,  Ludwig H.-G.,  Wedemeyer-B{\"o}hm S.,  Schaffenberger
  W.,   Steiner O.,  2012, Journal of Computational Physics, 231, 919

\bibitem[\protect\citeauthoryear{{Gaia Collaboration} et~al.,}{{Gaia
  Collaboration} et~al.}{2018}]{gaia2018}
{Gaia Collaboration} et~al., 2018, \mn@doi [\aap]
  {10.1051/0004-6361/201833051}, \href
  {http://adsabs.harvard.edu/abs/2018A%26A...616A...1G} {616, A1}

\bibitem[\protect\citeauthoryear{{G{\"a}nsicke}, {Schreiber}, {Toloza},
  {Fusillo}, {Koester}  \& {Manser}}{{G{\"a}nsicke} et~al.}{2019}]{pred2019}
{G{\"a}nsicke} B.~T.,  {Schreiber} M.~R.,  {Toloza} O.,  {Fusillo} N. P.~G.,
  {Koester} D.,   {Manser} C.~J.,  2019, \mn@doi [\nat]
  {10.1038/s41586-019-1789-8}, \href
  {https://ui.adsabs.harvard.edu/abs/2019Natur.576...61G} {576, 61}

\bibitem[\protect\citeauthoryear{{Genest-Beaulieu} \&
  {Bergeron}}{{Genest-Beaulieu} \& {Bergeron}}{2019a}]{genest2019a}
{Genest-Beaulieu} C.,  {Bergeron} P.,  2019a, arXiv e-prints, \href
  {http://adsabs.harvard.edu/abs/2019arXiv190101857G} {}

\bibitem[\protect\citeauthoryear{{Genest-Beaulieu} \&
  {Bergeron}}{{Genest-Beaulieu} \& {Bergeron}}{2019b}]{genest2019b}
{Genest-Beaulieu} C.,  {Bergeron} P.,  2019b, \mn@doi [\apj]
  {10.3847/1538-4357/ab379e}, \href
  {https://ui.adsabs.harvard.edu/abs/2019ApJ...882..106G} {882, 106}

\bibitem[\protect\citeauthoryear{{Gentile Fusillo}, {G{\"a}nsicke}, {Farihi},
  {Koester}, {Schreiber}  \& {Pala}}{{Gentile Fusillo}
  et~al.}{2017}]{gentilefusillo2017}
{Gentile Fusillo} N.~P.,  {G{\"a}nsicke} B.~T.,  {Farihi} J.,  {Koester} D.,
  {Schreiber} M.~R.,   {Pala} A.~F.,  2017, \mn@doi [\mnras]
  {10.1093/mnras/stx468}, \href
  {http://adsabs.harvard.edu/abs/2017MNRAS.468..971G} {468, 971}

\bibitem[\protect\citeauthoryear{{Gentile Fusillo} et~al.,}{{Gentile Fusillo}
  et~al.}{2019a}]{gentilefusillo2019}
{Gentile Fusillo} N.~P.,  et~al., 2019a, \mn@doi [\mnras]
  {10.1093/mnras/sty3016}, \href
  {http://adsabs.harvard.edu/abs/2019MNRAS.482.4570G} {482, 4570}

\bibitem[\protect\citeauthoryear{{Gentile Fusillo} et~al.,}{{Gentile Fusillo}
  et~al.}{2019b}]{gentilefusillo2019cat}
{Gentile Fusillo} N.~P.,  et~al., 2019b, VizieR Online Data Catalog, \href
  {http://adsabs.harvard.edu/abs/2019yCat..74824570G} {748}

\bibitem[\protect\citeauthoryear{{Gentile Fusillo}, {Tremblay}, {Bohlin},
  {Deustua}  \& {Kalirai}}{{Gentile Fusillo} et~al.}{2020}]{gentile2020}
{Gentile Fusillo} N.~P.,  {Tremblay} P.-E.,  {Bohlin} R.~C.,  {Deustua} S.~E.,
   {Kalirai} J.~S.,  2020, \mn@doi [\mnras] {10.1093/mnras/stz2984}, \href
  {https://ui.adsabs.harvard.edu/abs/2020MNRAS.491.3613G} {491, 3613}

\bibitem[\protect\citeauthoryear{{Gianninas}, {Bergeron}  \&
  {Ruiz}}{{Gianninas} et~al.}{2011}]{gianninas2011}
{Gianninas} A.,  {Bergeron} P.,   {Ruiz} M.~T.,  2011, \mn@doi [\apj]
  {10.1088/0004-637X/743/2/138}, \href
  {https://ui.adsabs.harvard.edu/abs/2011ApJ...743..138G} {743, 138}

\bibitem[\protect\citeauthoryear{{Grimm-Strele}, {Kupka}, {L{\"o}w-Baselli},
  {Mundprecht}, {Zaussinger}  \& {Schiansky}}{{Grimm-Strele}
  et~al.}{2015}]{Grimm_Strele_2pressurescale}
{Grimm-Strele} H.,  {Kupka} F.,  {L{\"o}w-Baselli} B.,  {Mundprecht} E.,
  {Zaussinger} F.,   {Schiansky} P.,  2015, \mn@doi [\na]
  {10.1016/j.newast.2013.11.005}, \href
  {http://adsabs.harvard.edu/abs/2015NewA...34..278G} {34, 278}

\bibitem[\protect\citeauthoryear{{Hermes}, {Kawaler}, {Bischoff-Kim},
  {Provencal}, {Dunlap}  \& {Clemens}}{{Hermes} et~al.}{2017}]{hermes2017}
{Hermes} J.~J.,  {Kawaler} S.~D.,  {Bischoff-Kim} A.,  {Provencal} J.~L.,
  {Dunlap} B.~H.,   {Clemens} J.~C.,  2017, \mn@doi [\apj]
  {10.3847/1538-4357/835/2/277}, \href
  {http://adsabs.harvard.edu/abs/2017ApJ...835..277H} {835, 277}

\bibitem[\protect\citeauthoryear{{Hummer} \& {Mihalas}}{{Hummer} \&
  {Mihalas}}{1988}]{hummer1988}
{Hummer} D.~G.,  {Mihalas} D.,  1988, \mn@doi [\apj] {10.1086/166600}, \href
  {http://adsabs.harvard.edu/abs/1988ApJ...331..794H} {331, 794}

\bibitem[\protect\citeauthoryear{{Iben}, {Kaler}, {Truran}  \&
  {Renzini}}{{Iben} et~al.}{1983}]{iben1983}
{Iben} I. J.,  {Kaler} J.~B.,  {Truran} J.~W.,   {Renzini} A.,  1983, \mn@doi
  [\apj] {10.1086/160631}, \href
  {https://ui.adsabs.harvard.edu/abs/1983ApJ...264..605I} {264, 605}

\bibitem[\protect\citeauthoryear{{Kalirai}}{{Kalirai}}{2012}]{kalirai2012}
{Kalirai} J.~S.,  2012, \mn@doi [\nat] {10.1038/nature11062}, \href
  {https://ui.adsabs.harvard.edu/abs/2012Natur.486...90K} {486, 90}

\bibitem[\protect\citeauthoryear{{Kepler} et~al.,}{{Kepler}
  et~al.}{2019}]{kepler2019}
{Kepler} S.~O.,  et~al., 2019, \mn@doi [\mnras] {10.1093/mnras/stz960}, \href
  {https://ui.adsabs.harvard.edu/abs/2019MNRAS.486.2169K} {486, 2169}

\bibitem[\protect\citeauthoryear{{Kilic}, {Bergeron}, {Dame}, {Hambly},
  {Rowell}  \& {Crawford}}{{Kilic} et~al.}{2019}]{kilic2019}
{Kilic} M.,  {Bergeron} P.,  {Dame} K.,  {Hambly} N.~C.,  {Rowell} N.,
  {Crawford} C.~L.,  2019, \mn@doi [\mnras] {10.1093/mnras/sty2755}, \href
  {https://ui.adsabs.harvard.edu/abs/2019MNRAS.482..965K} {482, 965}

\bibitem[\protect\citeauthoryear{{Klein} et~al.,}{{Klein}
  et~al.}{2020}]{klein2020}
{Klein} B.,  et~al., 2020, \mn@doi [\apj] {10.3847/1538-4357/ab9b24}, \href
  {https://ui.adsabs.harvard.edu/abs/2020ApJ...900....2K} {900, 2}

\bibitem[\protect\citeauthoryear{{Kleinman} et~al.,}{{Kleinman}
  et~al.}{2004}]{kleinman04}
{Kleinman} S.~J.,  et~al., 2004, \mn@doi [\apj] {10.1086/383464}, \href
  {https://ui.adsabs.harvard.edu/abs/2004ApJ...607..426K} {607, 426}

\bibitem[\protect\citeauthoryear{{Kleinman} et~al.,}{{Kleinman}
  et~al.}{2013}]{kleinman13}
{Kleinman} S.~J.,  et~al., 2013, \mn@doi [\apjs] {10.1088/0067-0049/204/1/5},
  \href {http://adsabs.harvard.edu/abs/2013ApJS..204....5K} {204, 5}

\bibitem[\protect\citeauthoryear{{Koester} \& {Kepler}}{{Koester} \&
  {Kepler}}{2015}]{koester2015}
{Koester} D.,  {Kepler} S.~O.,  2015, \mn@doi [\aap]
  {10.1051/0004-6361/201527169}, \href
  {http://adsabs.harvard.edu/abs/2015A%26A...583A..86K} {583, A86}

\bibitem[\protect\citeauthoryear{{Lindegren} et~al.,}{{Lindegren}
  et~al.}{2018}]{lindergren2018}
{Lindegren} L.,  et~al., 2018, \mn@doi [\aap] {10.1051/0004-6361/201832727},
  \href {https://ui.adsabs.harvard.edu/abs/2018A&A...616A...2L} {616, A2}

\bibitem[\protect\citeauthoryear{{Ludwig} \& {Steffen}}{{Ludwig} \&
  {Steffen}}{2008}]{ludwig08}
{Ludwig} H.-G.,  {Steffen} M.,  2008, in {Santos} N.~C.,  {Pasquini} L.,
  {Correia} A.~C.~M.,   {Romaniello} M.,  eds, Proceedings of Precision
  Spectroscopy in Astrophysics. ESO Astrophysics Symposia.
p.~133

\bibitem[\protect\citeauthoryear{{Ludwig}, {Jordan}  \& {Steffen}}{{Ludwig}
  et~al.}{1994}]{ludwig_1994_op_binning}
{Ludwig} H.-G.,  {Jordan} S.,   {Steffen} M.,  1994, \aap, \href
  {http://adsabs.harvard.edu/abs/1994A%26A...284..105L} {284, 105}

\bibitem[\protect\citeauthoryear{{Ludwig}, {Freytag}  \& {Steffen}}{{Ludwig}
  et~al.}{1999}]{ludwig1999}
{Ludwig} H.-G.,  {Freytag} B.,   {Steffen} M.,  1999, \aap, \href
  {http://adsabs.harvard.edu/abs/1999A%26A...346..111L} {346, 111}

\bibitem[\protect\citeauthoryear{{MacDonald} \& {Vennes}}{{MacDonald} \&
  {Vennes}}{1991}]{macdonald1991}
{MacDonald} J.,  {Vennes} S.,  1991, \mn@doi [\apj] {10.1086/169937}, \href
  {http://adsabs.harvard.edu/abs/1991ApJ...371..719M} {371, 719}

\bibitem[\protect\citeauthoryear{{Ma{\'\i}z Apell{\'a}niz} \&
  {Weiler}}{{Ma{\'\i}z Apell{\'a}niz} \& {Weiler}}{2018}]{jesus2018}
{Ma{\'\i}z Apell{\'a}niz} J.,  {Weiler} M.,  2018, \mn@doi [\aap]
  {10.1051/0004-6361/201834051}, \href
  {https://ui.adsabs.harvard.edu/abs/2018A&A...619A.180M} {619, A180}

\bibitem[\protect\citeauthoryear{{Manser} et~al.,}{{Manser}
  et~al.}{2019}]{manser2019}
{Manser} C.~J.,  et~al., 2019, \mn@doi [Science] {10.1126/science.aat5330},
  \href {https://ui.adsabs.harvard.edu/abs/2019Sci...364...66M} {364, 66}

\bibitem[\protect\citeauthoryear{{Narayan} et~al.,}{{Narayan}
  et~al.}{2019}]{narayan2019}
{Narayan} G.,  et~al., 2019, \mn@doi [\apjs] {10.3847/1538-4365/ab0557}, \href
  {https://ui.adsabs.harvard.edu/abs/2019ApJS..241...20N} {241, 20}

\bibitem[\protect\citeauthoryear{{Nordlund}}{{Nordlund}}{1982}]{nordlund_1982_opac_binning}
{Nordlund} A.,  1982, \aap, \href
  {http://adsabs.harvard.edu/abs/1982A%26A...107....1N} {107, 1}

\bibitem[\protect\citeauthoryear{{Provencal}, {Shipman}, {Riddle}  \&
  {Vuckovic}}{{Provencal} et~al.}{2003}]{provencal2003}
{Provencal} J.~L.,  {Shipman} H.~L.,  {Riddle} R.~L.,   {Vuckovic} M.,  2003,
  in {de Martino} D.,  {Silvotti} R.,  {Solheim} J.-E.,   {Kalytis} R.,  eds,
  NATO ASIB Proc. 105: White Dwarfs Vol. 105, NATO ASIB Proc. 105: White
  Dwarfs. p.~235

\bibitem[\protect\citeauthoryear{{Rolland}, {Bergeron}  \&
  {Fontaine}}{{Rolland} et~al.}{2018}]{rolland2018}
{Rolland} B.,  {Bergeron} P.,   {Fontaine} G.,  2018, \mn@doi [\apj]
  {10.3847/1538-4357/aab713}, \href
  {http://adsabs.harvard.edu/abs/2018ApJ...857...56R} {857, 56}

\bibitem[\protect\citeauthoryear{{Rolland}, {Bergeron}  \&
  {Fontaine}}{{Rolland} et~al.}{2020}]{rolland2020}
{Rolland} B.,  {Bergeron} P.,   {Fontaine} G.,  2020, \mn@doi [\apj]
  {10.3847/1538-4357/ab6602}, \href
  {https://ui.adsabs.harvard.edu/abs/2020ApJ...889...87R} {889, 87}

\bibitem[\protect\citeauthoryear{{Sbordone} et~al.,}{{Sbordone}
  et~al.}{2010}]{sbordone2010}
{Sbordone} L.,  et~al., 2010, \mn@doi [\aap] {10.1051/0004-6361/200913282},
  \href {https://ui.adsabs.harvard.edu/abs/2010A&A...522A..26S} {522, A26}

\bibitem[\protect\citeauthoryear{{Schatzman}}{{Schatzman}}{1948}]{schatzman48}
{Schatzman} E.,  1948, \mn@doi [\nat] {10.1038/161061b0}, \href
  {http://adsabs.harvard.edu/abs/1948Natur.161R..61S} {161, 61}

\bibitem[\protect\citeauthoryear{{Shipman}, {Provencal}, {Riddle}  \&
  {Vuckovic}}{{Shipman} et~al.}{2002}]{shipman2002}
{Shipman} H.~L.,  {Provencal} J.,  {Riddle} R.,   {Vuckovic} M.,  2002, in
  American Astronomical Society Meeting Abstracts \#200. p.~765

\bibitem[\protect\citeauthoryear{{Steffen}, {Ludwig}  \& {Freytag}}{{Steffen}
  et~al.}{1995}]{steffen1955}
{Steffen} M.,  {Ludwig} H.~G.,   {Freytag} B.,  1995, \aap, \href
  {https://ui.adsabs.harvard.edu/abs/1995A&A...300..473S} {300, 473}

\bibitem[\protect\citeauthoryear{{Straniero}, {Dom{\'\i}nguez}, {Imbriani}  \&
  {Piersanti}}{{Straniero} et~al.}{2003}]{straniero2003}
{Straniero} O.,  {Dom{\'\i}nguez} I.,  {Imbriani} G.,   {Piersanti} L.,  2003,
  \mn@doi [\apj] {10.1086/345427}, \href
  {https://ui.adsabs.harvard.edu/abs/2003ApJ...583..878S} {583, 878}

\bibitem[\protect\citeauthoryear{{Tassoul}, {Fontaine}  \& {Winget}}{{Tassoul}
  et~al.}{1990}]{tassoul1990}
{Tassoul} M.,  {Fontaine} G.,   {Winget} D.~E.,  1990, \mn@doi [\apjs]
  {10.1086/191420}, \href {http://adsabs.harvard.edu/abs/1990ApJS...72..335T}
  {72, 335}

\bibitem[\protect\citeauthoryear{{Tremblay}, {Bergeron}, {Kalirai}  \&
  {Gianninas}}{{Tremblay} et~al.}{2010}]{tremblay2010}
{Tremblay} P.~E.,  {Bergeron} P.,  {Kalirai} J.~S.,   {Gianninas} A.,  2010,
  \mn@doi [\apj] {10.1088/0004-637X/712/2/1345}, \href
  {https://ui.adsabs.harvard.edu/abs/2010ApJ...712.1345T} {712, 1345}

\bibitem[\protect\citeauthoryear{{Tremblay}, {Ludwig}, {Steffen}  \&
  {Freytag}}{{Tremblay} et~al.}{2013a}]{tremblay_2013_3dmodels}
{Tremblay} P.-E.,  {Ludwig} H.-G.,  {Steffen} M.,   {Freytag} B.,  2013a,
  \mn@doi [\aap] {10.1051/0004-6361/201220813}, \href
  {http://adsabs.harvard.edu/abs/2013A%26A...552A..13T} {552, A13}

\bibitem[\protect\citeauthoryear{{Tremblay}, {Ludwig}, {Steffen}  \&
  {Freytag}}{{Tremblay} et~al.}{2013b}]{tremblay_2013_spectra}
{Tremblay} P.-E.,  {Ludwig} H.-G.,  {Steffen} M.,   {Freytag} B.,  2013b,
  \mn@doi [\aap] {10.1051/0004-6361/201322318}, \href
  {http://adsabs.harvard.edu/abs/2013A%26A...559A.104T} {559, A104}

\bibitem[\protect\citeauthoryear{{Tremblay}, {Kalirai}, {Soderblom}, {Cignoni}
  \& {Cummings}}{{Tremblay} et~al.}{2014}]{tremblay2014}
{Tremblay} P.~E.,  {Kalirai} J.~S.,  {Soderblom} D.~R.,  {Cignoni} M.,
  {Cummings} J.,  2014, \mn@doi [\apj] {10.1088/0004-637X/791/2/92}, \href
  {https://ui.adsabs.harvard.edu/abs/2014ApJ...791...92T} {791, 92}

\bibitem[\protect\citeauthoryear{{Tremblay}, {Gianninas}, {Kilic}, {Ludwig},
  {Steffen}, {Freytag}  \& {Hermes}}{{Tremblay}
  et~al.}{2015}]{tremblay2015_elm}
{Tremblay} P.~E.,  {Gianninas} A.,  {Kilic} M.,  {Ludwig} H.~G.,  {Steffen} M.,
   {Freytag} B.,   {Hermes} J.~J.,  2015, \mn@doi [\apj]
  {10.1088/0004-637X/809/2/148}, \href
  {https://ui.adsabs.harvard.edu/abs/2015ApJ...809..148T} {809, 148}

\bibitem[\protect\citeauthoryear{{Tremblay}, {Cukanovaite}, {Gentile Fusillo},
  {Cunningham}  \& {Hollands}}{{Tremblay} et~al.}{2019}]{tremblay2019}
{Tremblay} P.-E.,  {Cukanovaite} E.,  {Gentile Fusillo} N.~P.,  {Cunningham}
  T.,   {Hollands} M.~A.,  2019, \mn@doi [\mnras] {10.1093/mnras/sty3067},
  \href {http://adsabs.harvard.edu/abs/2019MNRAS.482.5222T} {482, 5222}

\bibitem[\protect\citeauthoryear{{Tremblay} et~al.,}{{Tremblay}
  et~al.}{2020a}]{tremblay2020}
{Tremblay} P.~E.,  et~al., 2020a, \mn@doi [\mnras] {10.1093/mnras/staa1892},
  \href {https://ui.adsabs.harvard.edu/abs/2020MNRAS.tmp.2021T} {}

\bibitem[\protect\citeauthoryear{{Tremblay}, {Beauchamp}  \&
  {Bergeron}}{{Tremblay} et~al.}{2020b}]{patricktremblay2020}
{Tremblay} P.,  {Beauchamp} A.,   {Bergeron} P.,  2020b, arXiv e-prints, \href
  {https://ui.adsabs.harvard.edu/abs/2020arXiv200809834T} {p. arXiv:2008.09834}

\bibitem[\protect\citeauthoryear{{Uns{\"o}ld}}{{Uns{\"o}ld}}{1955}]{unsold_1955}
{Uns{\"o}ld} A.,  1955, {Physik der Sternatmospharen, MIT besonderer
  Berucksichtigung der Sonne.}

\bibitem[\protect\citeauthoryear{{Van Grootel}, {Fontaine}, {Brassard}  \&
  {Dupret}}{{Van Grootel} et~al.}{2017}]{vangrootel2017}
{Van Grootel} V.,  {Fontaine} G.,  {Brassard} P.,   {Dupret} M.-A.,  2017, in
  {Tremblay} P.-E.,  {Gaensicke} B.,   {Marsh} T.,  eds,  Astronomical Society
  of the Pacific Conference Series Vol. 509, 20th European White Dwarf
  Workshop. p.~321

\bibitem[\protect\citeauthoryear{{Vanderbosch} et~al.,}{{Vanderbosch}
  et~al.}{2019}]{vanderbosch2019}
{Vanderbosch} Z.,  et~al., 2019, arXiv e-prints, \href
  {https://ui.adsabs.harvard.edu/abs/2019arXiv190809839V} {p. arXiv:1908.09839}

\bibitem[\protect\citeauthoryear{{Vanderburg} et~al.,}{{Vanderburg}
  et~al.}{2015}]{vanderburg2015}
{Vanderburg} A.,  et~al., 2015, \mn@doi [\nat] {10.1038/nature15527}, \href
  {https://ui.adsabs.harvard.edu/abs/2015Natur.526..546V} {526, 546}

\bibitem[\protect\citeauthoryear{{Vanderburg} et~al.,}{{Vanderburg}
  et~al.}{2020}]{vanderburg2020}
{Vanderburg} A.,  et~al., 2020, \mn@doi [\nat] {10.1038/s41586-020-2713-y},
  \href {https://ui.adsabs.harvard.edu/abs/2020Natur.585..363V} {585, 363}

\bibitem[\protect\citeauthoryear{{Veras}}{{Veras}}{2016}]{veras2016}
{Veras} D.,  2016, \mn@doi [Royal Society Open Science] {10.1098/rsos.150571},
  \href {http://adsabs.harvard.edu/abs/2016RSOS....350571V} {3, 150571}

\bibitem[\protect\citeauthoryear{{Virtanen} et~al.,}{{Virtanen}
  et~al.}{2020}]{scipy2020}
{Virtanen} P.,  et~al., 2020, \mn@doi [Nature Methods]
  {https://doi.org/10.1038/s41592-019-0686-2}, \href {https://rdcu.be/b08Wh}
  {17, 261}

\bibitem[\protect\citeauthoryear{{V{\"o}gler}, {Bruls}  \&
  {Sch{\"u}ssler}}{{V{\"o}gler} et~al.}{2004}]{vogler_2004_op_binning}
{V{\"o}gler} A.,  {Bruls} J.~H.~M.~J.,   {Sch{\"u}ssler} M.,  2004, \mn@doi
  [\aap] {10.1051/0004-6361:20047043}, \href
  {http://adsabs.harvard.edu/abs/2004A%26A...421..741V} {421, 741}

\bibitem[\protect\citeauthoryear{{Werner} \& {Herwig}}{{Werner} \&
  {Herwig}}{2006}]{werner2006}
{Werner} K.,  {Herwig} F.,  2006, \mn@doi [\pasp] {10.1086/500443}, \href
  {https://ui.adsabs.harvard.edu/abs/2006PASP..118..183W} {118, 183}

\bibitem[\protect\citeauthoryear{{Wilson}, {G{\"a}nsicke}, {Koester}, {Toloza},
  {Pala}, {Breedt}  \& {Parsons}}{{Wilson} et~al.}{2015}]{wilson2015}
{Wilson} D.~J.,  {G{\"a}nsicke} B.~T.,  {Koester} D.,  {Toloza} O.,  {Pala}
  A.~F.,  {Breedt} E.,   {Parsons} S.~G.,  2015, \mn@doi [\mnras]
  {10.1093/mnras/stv1201}, \href
  {https://ui.adsabs.harvard.edu/abs/2015MNRAS.451.3237W} {451, 3237}

\bibitem[\protect\citeauthoryear{{Zuckerman}, {Koester}, {Melis}, {Hansen}  \&
  {Jura}}{{Zuckerman} et~al.}{2007}]{zuckerman2007}
{Zuckerman} B.,  {Koester} D.,  {Melis} C.,  {Hansen} B.~M.,   {Jura} M.,
  2007, \mn@doi [\apj] {10.1086/522223}, \href
  {https://ui.adsabs.harvard.edu/abs/2007ApJ...671..872Z} {671, 872}

\makeatother
\end{thebibliography}




\clearpage

\appendix

\section{Python 3D correction code} ~\label{sec:python_code}

The following Python code can be used to determine the 3D corrections for given 1D \y, \logg~and \teff~values. Brief description of how to use the code is also provided. The 3D corrections should only be applied to spectroscopically-determined 1D atmospheric parameters in the ranges $7.5 \leq$~\logg~$\leq 9.1$ dex, $11\,900 \leq$~\teff~$\leq 33\,900$ K and $-10.0 \leq$~\y~$\leq -2.0$ dex.

\newpage 
\begin{table*}
\begin{lstlisting}[language=Python]
import numpy as np
""" Correction functions:
corr_g -> log(g) correction function
corr_t -> Teff correction function """
def corr_g(x,u1,v1,w1):
    if u1 < 7.5 or u1 > 9.1 or v1 < 11900.0 or v1 > 33900 or w1 > -2.0:
        a = 0.*u1
    elif w1 < -10.0:
        w1=-10.0
    else:
        u = (u1 - 7.0)/7.0
        v = (v1 - 10000.0)/10000.0
        w = w1/(-10.)
        
        x = [1.03268078e-03, -4.05683312e-02, 2.22405930e-01, 6.51289937e+00,
            -3.73620338e+00, 1.55250184e+00, -2.38491665e+00, 8.54314405e-01,
            3.55696663e+00, -3.50421472e+00, -1.75128098e-02]
   
        a = (x[0] + x[1]*np.exp(x[2]+x[3]*u+x[4]*v+(x[5]+x[6]*np.exp(x[7]+
        x[8]*u+x[9]*v+x[10]*w))*w))
    return a 
def corr_t(x,u1,v1,w1):
    if u1 < 7.5 or u1 > 9.1 or v1 < 11900.0 or v1 > 33900 or w1 > -2.0 or w1 < -10.0:
        a = 0.*u1
    elif w1 < -10.0:
        w1=-10.0
    else:
        u = (u1 - 7.0)/7.0
        v = (v1 - 10000.0)/10000.0
        w = w1/(-10.)
        x = [-1.72633108e-03, 2.01885776e-02, -6.12179013e-01, -3.94213896e+00,
             3.00297254e+00, 1.97486482e-01, -3.98391235e+00, 5.17142871e+00,
             -7.52378709e+00, 3.78652321e+00, -4.76872681e+01, -3.88959969e-04,
             -2.19507125e+00, -6.95556318e+00, 1.41727246e+00, 4.76742530e+01]
    
        a = ( x[0] + x[1]*np.exp((x[2]+(x[6] + x[11]*np.exp(x[12]+x[13]*u+x[14]*v
        +x[15]*w))*np.exp(x[7]+x[8]*u+x[9]*v+x[10]*w))+x[3]*u+x[4]*v+x[5]*w))
    return a 
    
""" How to use:
For example, you want to find 3D log(g) correction for a 1D spectroscopically-determined values of 
log(H/He) = -2.3, log(g) = 8.45, Teff = 23890 K. Can also use lists of Teff, logg, logH/He."""

correction_in_logg = corr_g(8.45,23890,-2.3)

"""This will give the correction in log(g) which must be ADDED to 1D log(g) value in 
order to correct for 3D effects.
Same procedure can be repeated for 3D Teff and log(H/He) corrections."""

correction_in_teff = corr_t(8.45,23890,-2.3)

corrected_logg = 8.45 + 7.*correction_in_logg
corrected_teff = 23890 + 10000.*correction_in_teff


\end{lstlisting}
\end{table*}


\bsp	
\label{lastpage}
\end{document}